\documentclass[aps,twocolumn,showpacs,superscriptaddress,preprintnumbers,amsmath,amssymb,pra,10pt]{revtex4-1}
\usepackage{graphicx}
\usepackage{dcolumn}
\usepackage{appendix}
\usepackage{bm}

\usepackage[usenames,dvipsnames]{xcolor}
\usepackage{subfigure}
\usepackage{bbm}
\usepackage{enumerate}
\usepackage[
  bookmarks=true,
  colorlinks,
  linkcolor=blue,
  urlcolor=blue,
  citecolor=blue,
  plainpages=false,
  pdfpagelabels,
  final,
  breaklinks=true
]{hyperref}

\usepackage{titlesec}
\usepackage{array}
\usepackage{dsfont}
\usepackage{physics}

\newcommand{\relaxket}[1]{\lvert{#1}\rangle}
\newcommand{\relaxbra}[1]{\langle{#1}\rvert}

\begin{document}

\title{Weak measurement in strong laser field physics}

\author{Philipp Stammer}
\email{philipp.stammer@icfo.eu}
\affiliation{ICFO -- Institut de Ciencies Fotoniques, The Barcelona Institute of Science and Technology, 08860 Castelldefels (Barcelona), Spain}
\affiliation{Atominstitut, Technische Universität Wien, Stadionallee 2, 1020 Vienna, Austria}

\author{Javier Rivera-Dean}
\affiliation{ICFO -- Institut de Ciencies Fotoniques, The Barcelona Institute of Science and Technology, 08860 Castelldefels (Barcelona), Spain}

\author{Marcelo~F.~Ciappina}
\affiliation{Department of Physics, Guangdong Technion - Israel Institute of Technology, 241 Daxue Road, Shantou, Guangdong, 515063, China}
\affiliation{Technion - Israel Institute of Technology, Haifa, 32000, Israel}
\affiliation{Guangdong Provincial Key Laboratory of Materials and Technologies for Energy Conversion, Guangdong Technion - Israel Institute of Technology, 241 Daxue Road, Shantou, Guangdong, 515063, China}

\author{Maciej Lewenstein}
\affiliation{ICFO -- Institut de Ciencies Fotoniques, The Barcelona Institute of Science and Technology, 08860 Castelldefels (Barcelona), Spain}
\affiliation{ICREA, Pg. Llu\'{\i}s Companys 23, 08010 Barcelona, Spain}

\date{\today}

\begin{abstract}

The advantage of attosecond measurements is the possibility of time-resolving ultrafast quantum phenomena of electron dynamics. Many such measurements are of interferometric nature, and therefore give access to the phase. Likewise, weak measurements are intrinsically interferometric and specifically take advantage of interfering probability amplitudes, therefore encoding the phase information of the process. In this work, we show that attosecond interferometry experiments can be seen as a weak measurement, which unveils how this notion is connected to strong field physics and attosecond science. In particular, we show how the electron trajectory picks up a new phase, which occurs due to the weak measurement of the process. This phase can show significant contributions in the presence of spectral features of the measured system. Furthermore, extending this approach to include non-classical driving fields shows that the generated harmonics exhibit non-trivial features in their quantum state and photon statistics. This opens the path towards investigations of \textit{attosecond quantum interferometry} experiments.

\end{abstract}

\maketitle

\section{\label{sec:intro}Introduction}

Strong laser field physics investigates the interaction of intense laser fields with matter, leading to a highly non-linear response of the material. Prominent processes concern field induced optical tunneling and strong field ionization~\cite{keldysh2024ionization, ivanov2005anatomy}, or the process of light generation via high-order harmonic generation (HHG)~\cite{lewenstein1994theory}, which are at the heart of attosecond physics~\cite{krausz2009attosecond}. While most of these intense laser driven processes have been described by semi-classical methods for decades~\cite{amini2019symphony}, there is a recent and growing interest in studying the phenomena from a quantum optical perspective~\cite{cruz2024quantum, stammer2025theory}. This includes the study of entanglement of the emitted radiation~\cite{stammer2022high, stammer2024entanglement, yi2024generation, theidel2024evidence}, between light and matter~\cite{rivera2022light, stenquist2025entanglement} or within the ionization process itself~\cite{maxwell2022entanglement, ruberti2024bell, stenquist2025entanglement}. 

However, this work focuses on a new and thus far unexplored direction on quantum optical strong field physics, namely, studying the scheme of attosecond interferometry techniques from a quantum information perspective. In doing so, we introduce concepts from the quantum theory of measurement to the realm of intense laser driven systems~\cite{wiseman2009quantum, stammer2022theory}. In particular, we investigate the process of HHG driven by an intense laser field of frequency $\omega$ and its perturbative second harmonic. Due to the broken temporal symmetry by the presence of the perturbative $2\omega$ field, the emitted harmonics show additional peaks at even harmonic frequencies~\cite{gaarde1996theory, dahlstrom2011quantum}, whereas the conventional single color HHG only emits odd harmonics for inversion symmetric targets~\cite{lewenstein1994theory}. 
We show that the presence of the perturbative field can be understood as a weak measurement (WM) process~\cite{aharonov1988result, dressel2014colloquium}, in which the intense laser driven electron acquires an additional phase shift due to the WM. 
In this scheme, the weak perturbative field allows information to be extracted with minimal disturbance, and the interferometric nature of the attosecond measurements allows to get access to this phase. 
This close relation between attosecond experiments and WM has so far remained unnoticed, although the notion of WM makes use of interfering probability amplitudes~\cite{duck1989sense}. With this it gives experimental access to the phase information of the process. Since many strong field phenomena and attosecond experiments make particular use of interference, the concept of WM provides a general framework for analyzing such experiments, 
and a direct way of extracting the encoded phase underlying these dynamics.

\begin{figure}
    \centering
	\includegraphics[width = 1.0\columnwidth]{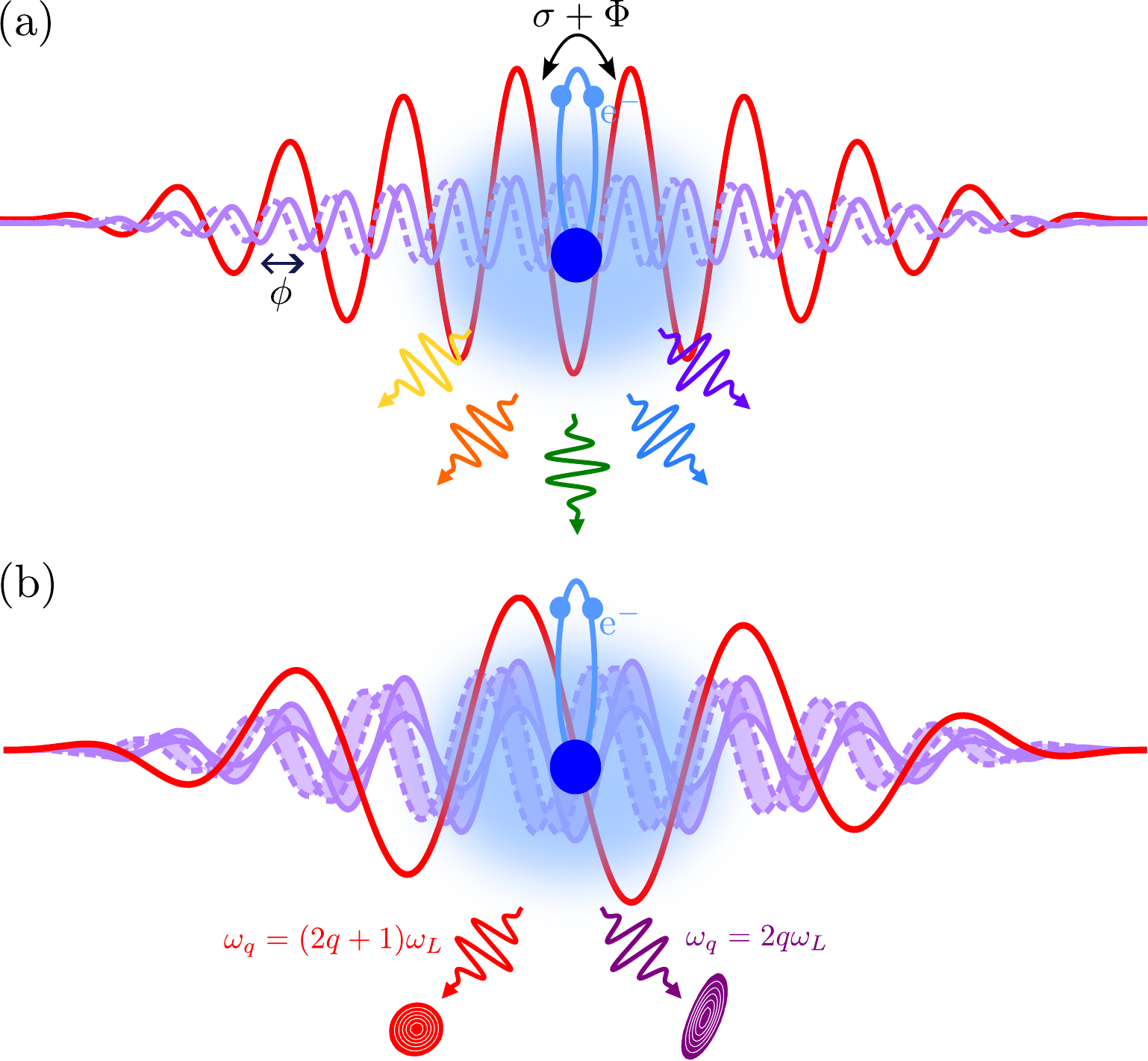}
	\caption{\textbf {Attosecond quantum interferometry} A strong laser field $\omega$ and its weak second harmonic $2\omega$ field drive the process of HHG in atoms with a variable phase difference $\phi$. Due to the presence of the perturbative $2\omega$ field the electron acquires a phase $\sigma$ due to the propagation in the continuum and a new phase shift $\Phi$ arising from a WM perspective. In (a) the atom is driven by classical coherent fields, while in (b) the perturbative second harmonic light shows squeezing in the field quadrature. Due to the presence of the squeezing the generated even ($\omega_q = 2 q \omega$) and odd ($\omega_q = (2q+1)\omega$) harmonics show different characteristics in their Wigner functions.   }
      \label{fig:scheme}
\end{figure}

In the context of attosecond processes, weak measurements arise when two interactions at two different times can lead to the same final observation, such that the two pathways can not be distinguished, and therefore the probability amplitudes interfere. The two interactions at different times can be understood as the analog of two spatial slits in a typical double-slit experiment. This temporal double-slit experiment is realized when strong field ionization is considered for a full cycle pulse, such that the ionization bursts at each half cycle lead to the same final state and interfere (see Fig.~\ref{fig:scheme}). This is precisely the scenario in HHG, and particularly attosecond interferometry experiments~\cite{pedatzur2015attosecond, dahlstrom2011quantum, dudovich2006measuring}.

Combining strong laser field physics with weak measurements, places the work along the rapidly growing field of quantum optical investigations of HHG, which has so far mainly focused on finding the quantum state of the harmonic field modes~\cite{lewenstein2021generation, gorlach2020quantum, stammer2023quantum}. This has shown interesting insights into the field properties such as squeezing~\cite{stammer2024entanglement, yi2024generation, tzur2024generation, theidel2024observation, lange2024electron}, non-classical Wigner functions of the driving field~\cite{lewenstein2021generation, rivera2022strong}, or the presence of classical~\cite{lemieux2025photon} and non-classical photon statistics~\cite{stammer2025theory}. 
However, here we extend such studies towards a quantum optical version of attosecond interferometry experiments, by considering non-classical driving fields and study the quantum state of the generated harmonics. We call this approach \textit{attosecond quantum interferometry (AQI)}. With this we can generate non-trivial quantum state of the harmonic modes showing quadrature stretching, and find non-classical photon statistics in the emission characteristics of the harmonics.  
With this work, we leverage the quantum optical studies to include concepts from quantum theory of measurements towards connecting strong field physics with quantum information science.

\section{Results}

\subsection*{Weak measurement in strongly driven atoms}

First, we introduce the quantum optical framework for strongly driven atomic systems by an intense laser field of frequency $\omega$ and a second harmonic perturbative field $2\omega$. The formal approach closely follows the single-color quantum optical description of the HHG process in atoms introduced in Ref.~\cite{lewenstein2021generation}. The scheme of HHG driven by a strong fundamental field together with its perturbative second harmonic was studied in the classical regime~\cite{pedatzur2015attosecond, dahlstrom2011quantum}, which we now extend by providing the quantum optical counterpart of attosecond interferometry.
We start by considering the quantum state of the initial field, which is given by a product of coherent states $\ket{\Phi(0)} = \ket{\alpha_1}\otimes \ket{\alpha_2} \otimes \ket{\{ 0_q \}}$, where the driving laser and its second harmonic are given by the coherent states $\ket{\alpha_1}$ and $\ket{\alpha_2}$, respectively, and the harmonic field modes $q \ge 3$ are initially in the vacuum $\ket{\{ 0_q\}} = \bigotimes_{q\ge 3} \ket{0_q}$. 
We aim to solve the Schrödinger equation for the field modes
\begin{align}
\label{eq:schrödinger}
    i \hbar \partial_t \ket{\Phi(t)} = H(t) \ket{\Phi(t)},
\end{align}
taking into account the light emission of a strongly driven atom leading to HHG~\cite{lewenstein2021generation, stammer2023quantum}.
For typical HHG experiments in gases, such that the emitters are uncorrelated~\cite{sundaram1990high} and depletion effects of the ground state are negligible~\cite{stammer2024entanglement}, the solution of the Schrödinger equation is known~\cite{lewenstein2021generation}. 
In this regime, the effective interaction Hamiltonian governing the evolution of the light field is given by $H(t) = - \expval{\vb{d}(t)} \cdot \vb{E}_Q(t)$, where $\expval{\vb{d}(t)} = \bra{g} \vb{d}(t) \ket{g}$ is the time-dependent dipole moment of the electron driven by the classical laser field starting in the ground state $\ket{g}$, and is coupled to the electric field operator $\vb{E}_Q(t) = - i f(t) \sum_{q=1}^{q_c} \kappa_q \vb{\epsilon}_q \left( b_q^\dagger e^{i \omega_q t} - b_q e^{- i \omega_q t} \right)$, where $\kappa_q$ is the light-matter coupling, $\vb{\epsilon}_q$ is the unit polarization vector and $b_q^{(\dagger)}$ is the annihilation (creation) operator in mode $q$ of frequency $\omega_q = q \omega$. The ultrashort nature of the pulse is taken into account via an envelope function $f(t)$.  
The solution to the Schrödinger equation in the low depletion limit~\eqref{eq:schrödinger} is given by a multi-mode coherent product state~\cite{lewenstein2021generation};
\begin{align}\label{Eq:QO:state:coh}
    \ket{\Phi(t)} = \ket{\alpha_1 + \chi_1} \otimes \ket{\alpha_2 + \chi_2} \otimes \ket{\{ \chi_q \}},
\end{align}
where the amplitudes are given by the Fourier transform of the time-dependent dipole moment $\chi_q = \kappa_q \int_{- \infty}^{\infty} dt f(t) \expval{\vb{d}(t)} e^{i q \omega t}$.    
The time-dependent dipole $\expval{\vb{d}(t)}$ is driven by the classical coherent field $\vb{E}_{cl}(t) = \Tr[\vb{E}_Q(t) \dyad{\Phi(0)}]$, given by
\begin{align}
\label{eq:field_classical}
    \vb{E}_{cl}(t)  = 2 \kappa f(t) \left[ \abs{\alpha_1} \sin(\omega t) + \abs{\alpha_2} \sin(2\omega t + \phi) \right],
\end{align}
where $\phi$ is the relative phase between the two fields, and can therefore be solved by conventional methods in strong field physics~\cite{amini2019symphony}.
Now, from the amplitudes $\chi_q$ we can consider an instrumental example in which only a single cycle of the pulse, with time duration $T=2\pi / \omega$, is taken into account. For a vanishing $2\omega$ field, $\alpha_2 = 0$, it is well known that the interference between two consecutive half cycles leads to vanishing even harmonics, when using that $\expval{\vb{d}(t+T/2)} = - \expval{\vb{d}(t)}$ for inversion symmetric matter systems~\cite{dahlstrom2011quantum}.
In contrast, the presence of the second harmonic field breaks this symmetry, leading to the presence of even and odd harmonics~\cite{gaarde1996theory, pedatzur2015attosecond, dahlstrom2011quantum}. 
However, to obtain the harmonic amplitudes one needs to solve the time-dependent dipole moment $\expval{\vb{d}(t)}$ dressed by the classical fields. 
Considering that the driving field consists of a strong fundamental laser and its perturbative second harmonic ($\abs{\alpha_2} \ll \abs{\alpha_1}$), the dipole moment acquires additional components compared to the unperturbed case. 
Solving the dynamics of the perturbed dipole moment allows to obtain new correction terms to its phase (see Methods).
The correction term to the action is given by 
\begin{align}
    \Delta S( t, t_1,\vb{p}, \phi) = \sigma( t,t_1, \vb{p}, \phi) + \operatorname{Re}[\Phi( t,t_1, \vb{p}, \phi)].
\end{align}

The correction to the phase due to the $2\omega$ perturbation comes from an energy correction in the action during the propagation in the continuum $\sigma( t,t_1, \vb{p}, \phi)$, and a second phase term $\Phi ( t,t_1, \vb{p}, \phi)$.
One of the main new results in the WM scenario considered here, is an additional phase contribution from the ionization and recombination matrix elements
\begin{align}
\label{eq:correation_phase}
    \Phi ( t,t_1, \vb{p}, \phi) = A_{2\omega}(t_1, \phi) D_{ji}(t_1) - A_{2\omega} (t, \phi) D_{ij}(t).
\end{align}

Furthermore, and highlighting the connection of attosecond interferometry and the idea of WM, the crucial factor is written as 
\begin{align}
\label{eq:weak_value_general}
    D_{ij}(t) =   \frac{\bra{g}   d_i d_j   \ket{\vb{p} + \vb{A}(t)}  }{\bra{g} d_i \ket{\vb{p} + \vb{A}(t)}}, 
\end{align}
where $d_j = \boldsymbol{\epsilon}_2 \cdot \vb{d} $ with $\boldsymbol{\epsilon}_{2}$ the polarization unit vector of the $2\omega$ field.
This is the weak value (WV) of the dipole moment for the transition matrix element, and can in general be complex-valued~\cite{dressel2014colloquium, dressel2015weak}.

In addition to the correction of the phase, there is an exponential correction term $\Delta F(t_1, \phi)$ for the tunnel amplitude, originated from the ionization matrix element
\begin{align}
    \Delta F (t_1,\phi) = \frac{F_{2\omega}(t_1,\phi)}{F_{\omega}(t_1)} = \epsilon \frac{\sin(2\omega t_1 + \phi)}{\sin(\omega t_1)}.
\end{align}

This takes into account that the tunneling barrier is modified due to the presence of the perturbative field, which can either increase or decrease its width by varying the relative phase $\phi$.
While this factor appears from the ionization matrix element only (see Methods), and takes into account the deformed barrier for tunnel ionization, the additional phase factor $\Phi$ influences the entire electron dynamics.

To summarize the first main finding of this work, we have derived two new contributions that appear in the dipole moment corrections. These are the phase contribution $\Phi$, and the barrier suppression term $\Delta F$.
These terms have thus far been omitted, and the only attention has been given to the energy shift $\sigma( t,t_1, \vb{p}, \phi)$, whose consequences on HHG have been already studied in Refs.~\cite{dahlstrom2011quantum, pedatzur2015attosecond} to obtain information about the ultrafast electron dynamics.

However, in the traditional Lewenstein model~\cite{lewenstein1994theory} the phase of the transition dipole moment is assumed to vary slowly, and is thus not taken into account in the evaluation of the integrals in \eqref{eq:dipole_SFA_perturbed} via the saddle point method. This is not possible in systems with a rapidly varying transition moment phase, and the influence on the all-optical attosecond measurements of this phase has been investigated in Ref.~\cite{brown2022attosecond}. 
Here, in contrast, we have additional contributions to the phase from the first order correction of the $2\omega$ perturbation within the considered WM scenario. Thus, the particular case of a rapidly changing transition moment cross section, e.g. Cooper minima in Argon \cite{brown2022attosecond}, can still be added if necessary, for the atomic species under investigation. We thus go beyond existing work since the phase $\Phi(t, t_1, \vb{p}, \phi)$ also appears in cases where the transition dipole moment itself is slowly varying. 

With the WM scheme in attosecond interferometry, we can now consider the following questions. How can the real and imaginary parts of the WV be extracted experimentally, and how do they influence the electron trajectory? Furthermore, within the quantum optical framework used in this work, we can look at the backaction of the perturbation on the field itself \cite{lewenstein2021generation}, and study if this allows for quantum state engineering of light \cite{stammer2022theory, stammer2023quantum}. Answering those questions will be discussed in the remaining part of this paper. 

\subsection*{Perturbed quantum trajectories}

We have previously seen that the proper phase factor in~\eqref{eq:phase_total}, up to first order in the perturbation, includes three additional terms compared to the unperturbed case. One for each step in the celebrated 3-step model. We can thus ask about the influence of these phase factors to the quantum trajectories associated to the saddle point solutions of the integrals of the Fourier transform of the perturbed dipole moment $\expval{\vb{d} (q \omega)} = \int dt \expval{\vb{d}'(t)} e^{i q \omega t}$.

The saddle point equations for the integrals over $\{t_1, \vb{p}, t\}$, which define the saddle points $\{ t_i, \vb{p}_s, t_r \}$, are respectively given by  
\begin{align}
\label{eq:saddle_point_equations_ionization}
     \frac{ \left[ \vb{p}_s  + \vb{A}(t_i) \right]^2}{2} + I_p & =  -\Delta I_p(t_i, \vb{p}_s, t_r) \\
\label{eq:saddle_point_equations_propagation}
     \int_{t_i}^{t_r} d\tau \left[ \vb{p}_s + \vb{A}(\tau) \right] & = - \Delta x(t_i, \vb{p}_s, t_r) \\
\label{eq:saddle_point_equations_recombination}
     \frac{ \left[ \vb{p}_s + \vb{A}(t_r) \right]^2}{2} + I_p - q \omega & = - \Delta E_{kin} (t_i, \vb{p}_s, t_r)
\end{align}

\begin{figure}
    \centering
	\includegraphics[width = 1\columnwidth]{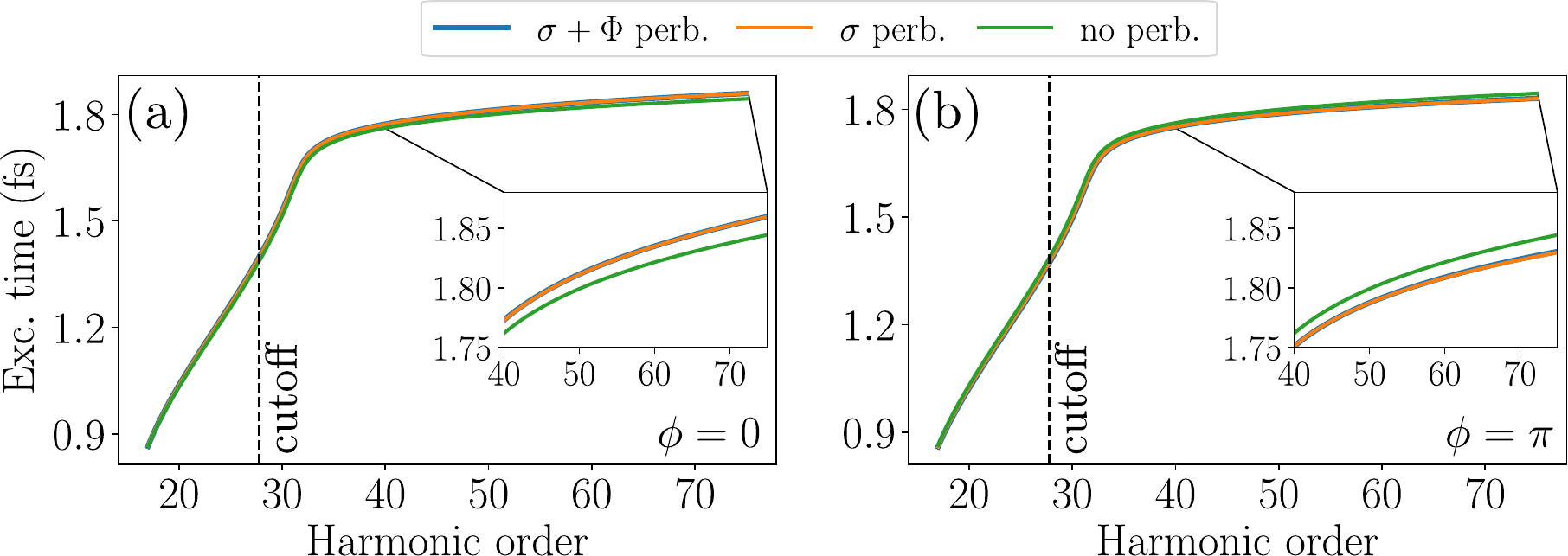}
	\caption{\textbf {Electron trajectory} Excursion time of the electron during the propagation in the continuum between the ionization time $t_i$ and recombination time $t_r$ from the perturbed saddle point equations of Eqs.~\eqref{eq:saddle_point_equations_ionization}-\eqref{eq:saddle_point_equations_recombination} for different harmonic orders. While the excursion time of the electron is mainly unchanged by the perturbative $2\omega$ field, the phase difference of $\phi = 0$ in (a) and $\phi = \pi$ in (b) either retards or advances the electron propagation time, respectively. 
    }
      \label{fig:Exc:time}
\end{figure}

Without taking into account the perturbations of the $2\omega$ field in the saddle point analysis, the right-hand side of the saddle point equations would vanish, and recover the conventional saddle point equations for the HHG process~\cite{lewenstein1994theory, amini2019symphony}. 
The influence of the additional phase terms on the saddle point equations can be seen as (i) for the ionization process the additional terms in \eqref{eq:saddle_point_equations_ionization} act as a change in the ionization potential, leading to an effective ionization potential $I_{p,eff} = I_p + \Delta I_p$, where
\begin{equation}
    \Delta I_p(t_i, \vb{p}_s, t_r) = [\vb{p}_s + \vb{A}(t_i)]\cdot \vb{A}_{2\omega}(t_i) - \partial_{t_i} \operatorname{Re} \Phi, 
\end{equation}
which can increase or decrease the ionization potential by changing the phase $\phi$. (ii) For the propagation in \eqref{eq:saddle_point_equations_propagation} the presence of the $2\omega$ field leads to an additional contribution of the overall displacement in the continuum
\begin{equation}
    \Delta x (t_i, \vb{p}_s, t_r) = \int_{t_i}^{t_r} d\tau \vb{A}_{2\omega}(\tau) + \nabla_{\vb{p}_s} \operatorname{Re} \Phi,
\end{equation}
and (iii) adds a correction to the kinetic energy at the recombination time for the energy conservation relation~\cite{smirnova2014multielectron}, where 
\begin{equation}
    \Delta E_{kin}(t_i, \vb{p}_s, t_r) = \left[ \vb{p}_s + \vb{A}(t_r) \right] \cdot \vb{A}_{2\omega} (t_r) + \partial_{t_r} \operatorname{Re} \Phi.
\end{equation}

From a physical picture, this $\omega-2\omega$ scheme manipulates the thickness of the tunneling barrier on a sub-cycle time scale, resulting in a complex phase shift between the two amplitudes of each ionization event in consecutive half cycles. This phase shift will be imprinted in the interference pattern of the temporal double-slit, i.e. the HHG spectrum. 
With the temporal analog of the double-slit we essentially probe the phase difference from tunneling through slightly different potential barriers.
The real part of this additional phase will cause a shift of the peak position whereas the imaginary part result in a decay within the barrier. Since the considered interference occurs in the temporal domain, we essentially probe the phase shift from tunneling through different potential barriers. This complex phase shift is caused by the WM of the electron dynamics. 
We have thus probed the electron wavefunction as it propagates under the field induced tunneling barrier by means of a WM~\cite{lundeen2011direct}. This was possible since the WM does not destroy the coherence properties of the investigated process, and thus the coherent electron wavepacket in the continuum conserves the memory about the tunneling process. The characteristics of the tunneling process and subsequent dynamics are imprinted in the emitted high harmonic radiation.
We note that the total phase information including the correction terms, and therefore the WV of the transition dipole moment in Eq.~\eqref{eq:weak_value_general}, can be measured by standard attosecond interferometry techniques, which correspond to the normalized sum or difference of an adjacent pair of even and odd harmonics~\cite{pedatzur2015attosecond}.

To get a first idea on the influence of the correction terms, we show in Figure~\ref{fig:Exc:time} the excursion time of the electron as a function of the harmonic order, computed for different scenarios: using the saddle-point equations for a single-color driver (green), including the effect of the perturbation $\sigma$ (orange), and accounting for both $\sigma$ and the WV contribution $\Phi$ (blue). As observed, the inclusion of $\sigma$ introduces moderate modifications to the excursion time. 
Different two-color phase delays $\phi$ leads to retarded ($\phi =0$) or advanced ($\phi = \pi$) excursion times of the electron. However, in both cases, the contribution of $\Phi$ remains negligible at the level of the electron excursion time.

\begin{figure}
    \centering
	\includegraphics[width = 1\columnwidth]{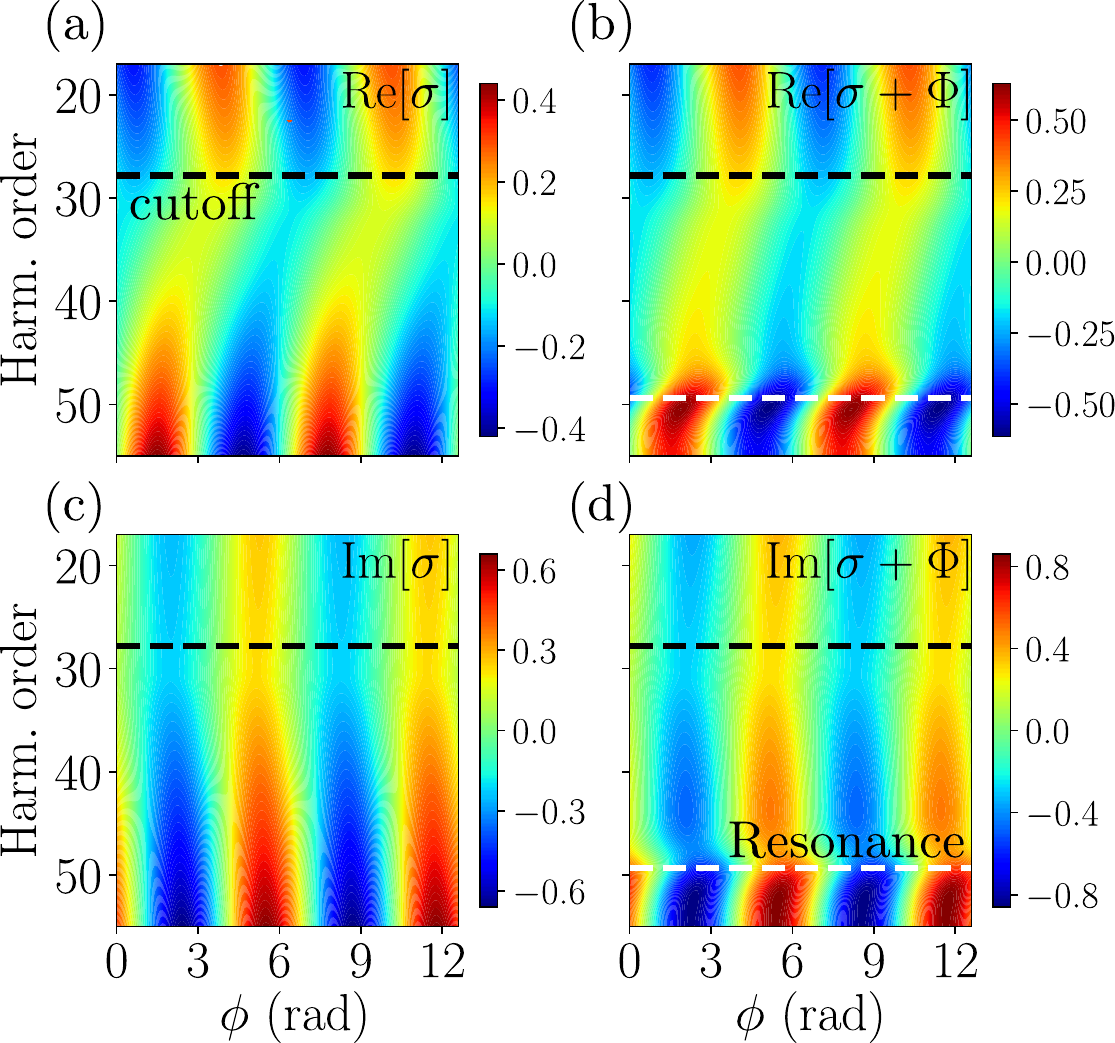}
	\caption{\textbf {HHG spectrum including a Fano resonance} Including the presence of a Fano resonance in the dipole transition matrix element as in Eq.~\eqref{eq:dipole_Fano} shows distinct features in the perturbation phase when including the new phase from the weak measurement scenario $\Phi$. Here, we have set $E_{\omega} = 0.053$ a.u., $E_{2\omega} = 10^{-2} E_{\omega}$, $\omega = 0.057$ a.u., and $I_p = 0.9$ a.u. corresponding to Helium.
    }
      \label{fig:Spec:Fano}
\end{figure}

This situation changes considerably in the presence of spectral features, such as autoionizing resonances, where we find that the WV of the dipole moment can significantly influence the spectral properties obtained in attosecond interferometry experiments. To account for such a resonance, we generalize the transition dipole matrix elements as in Refs.~\cite{rzazewski1983photoexcitation,lewenstein1983photon};
\begin{equation}
\label{eq:dipole_Fano}
	d_{\text{F},i}(v)
		= \dfrac{d_i(v)}{\sqrt{4\pi \Gamma}}
			\dfrac{\Gamma}{v^2/2 - \omega_R - i\Gamma}
			- \dfrac{i}{1-i q},
\end{equation}
where $d_i(v) = \langle g \vert \hat{d}_i \vert v\rangle$, $\Gamma$ denotes the inverse lifetime of the autoionizing state, $q$ is the Fano asymmetry parameter, and $\omega_R$ is the resonance frequency. Here, we set these quantities to $\Gamma = 0.2$, $q=1$ and $\omega = 1.913$ a.u., which suffice to introduce distinctive spectral signatures while keeping the weak value as a perturbative correction to the dipole moment (see Supplementary Material).

Figure~\ref{fig:Spec:Fano} illustrates the real and imaginary parts of $\sigma$ [Fig.~\ref{fig:Spec:Fano}~(a) and (c)], without the WV contribution, and of $\sigma + \Phi$ [Fig.~\ref{fig:Spec:Fano}~(b) and (d)], with the WV included. As shown in Ref.~\cite{pedatzur2015attosecond}, these quantities can be experimentally extracted by comparing the intensities of adjacent even and odd harmonic orders. Here, we observe that they become particularly sensitive to the WV in the presence of an autoionizing resonance. Specifically, while at low harmonic orders the influence of the resonance appears negligible, near the resonance frequency (around the 43rd harmonic order), the spectral response exhibits a pronounced enhancement that vanishes in the absence of the autoresonance. This result highlights attosecond tunneling interferometry as a powerful probe of weak values associated with the dipole moment.

\section*{Quantum optical attosecond interferometry}

In the results presented thus far, we have employed combinations of classical coherent $\omega-2\omega$ fields to demonstrate the sensitivity of attosecond interferometry to the WV of the dipole moment. In the following sections we shift our focus to a related but distinct question: How sensitive are attosecond interferometry setups to non-classical driving light fields, and can they be used to control the quantum state of the generated harmonics? To explore these questions, we now consider the scenario where the weak $2\omega$ field component exhibits quantum characteristics, realized by preparing it in a displaced squeezed vacuum (DSV) state [Fig.~1~(b)]. Accordingly, the initial quantum optical state is taken as $\ket{\Phi(0)} = \ket{\alpha_1}\otimes [\hat{D}_{2}(\alpha_2)\hat{S}(\xi)\ket{0}]\otimes \ket{\{0_q\}}$, where $\hat{D}_2(\alpha) = \text{exp}[\alpha\hat{a}_2^\dagger - \alpha^* \hat{a}_2]$ and $\hat{S}_2(\xi) = \text{exp}[\xi^* \hat{a}_2^2-\xi \hat{a}_2^{\dagger 2}]$ are the displacement and squeezing operator, respectively.

To describe the light-matter interaction dynamics under the presence of squeezed light and in the low-depletion regime~\cite{lewenstein2021generation,rivera2022strong,stammer2023quantum,stammer2024entanglement}, it is particularly convenient to represent the initial state of the $2\omega$ field component using the generalized positive $P$-representation~\cite{drummond_generalised_1980};
\begin{equation}
	\hat{\rho}_2(0)
		= \int \dd^2\alpha\int \dd^2 \beta \dfrac{P(\alpha,\beta^*)}{\braket{\beta^*}{\alpha}}
				\dyad{\alpha}{\beta^*},
\end{equation}
where $P(\alpha,\beta^*)$ is chosen to be a positive-definite function encoding the quantum statistical properties of the $2\omega$ mode~\cite{d_drummond_quantum_2016}. Accordingly, we write the initial quantum optical state as $\hat{\rho}(0) = \dyad{\Phi(0)} = \dyad{\alpha_1}\otimes \hat{\rho}_2(0) \otimes \dyad{\{0_q\}}$.

Under these conditions, we consider scenarios where the amount of squeezing is fixed to a specific value $\xi$, while the two-color phase difference $\phi$ of the coherent state amplitude varies as $\alpha_{2\omega} = \abs{\alpha_{2\omega}}e^{i\phi}$. This introduces a key distinction from conventional attosecond interferometry setups, where variations in $\phi$ merely control the relative phase between the $\omega$ and $2\omega$ drivers. In contrast, the present configuration allows for a continuous modulation of the quantum statistical nature of the $2\omega$ component, effectively sweeping from phase squeezing ($\phi = 0$) to amplitude squeezing ($\phi = \pi/2$). This allows to realize attosecond quantum interferometry setups. Given that the total intensity of the DSV state is given by $I_{2\omega} \propto \abs{\alpha_{2\omega}}^2 + \sinh[2](\abs{\xi}) \equiv I^{(\text{coh})}_{2\omega} + I^{(\text{squ})}_{2\omega}$, we work in a regime where $I_{2\omega} \ll I^{(\text{coh})}_{\omega}$, while imposing $I^{(\text{coh})}_{2\omega} \approx I^{(\text{squ})}_{2\omega}$.~That is, although the total $2\omega$-field intensity remains perturbative compared to the classical $\omega$ field, the squeezed and coherent components contribute comparably. In practice, this corresponds to $I_{2\omega} \sim 10^{10}-10^{11}$ W/cm$^2$, which are within reach of current state-of-art capabilities~\cite{rasputnyi2024high}. 

\begin{figure}
    \centering
    \includegraphics[width=1\columnwidth]{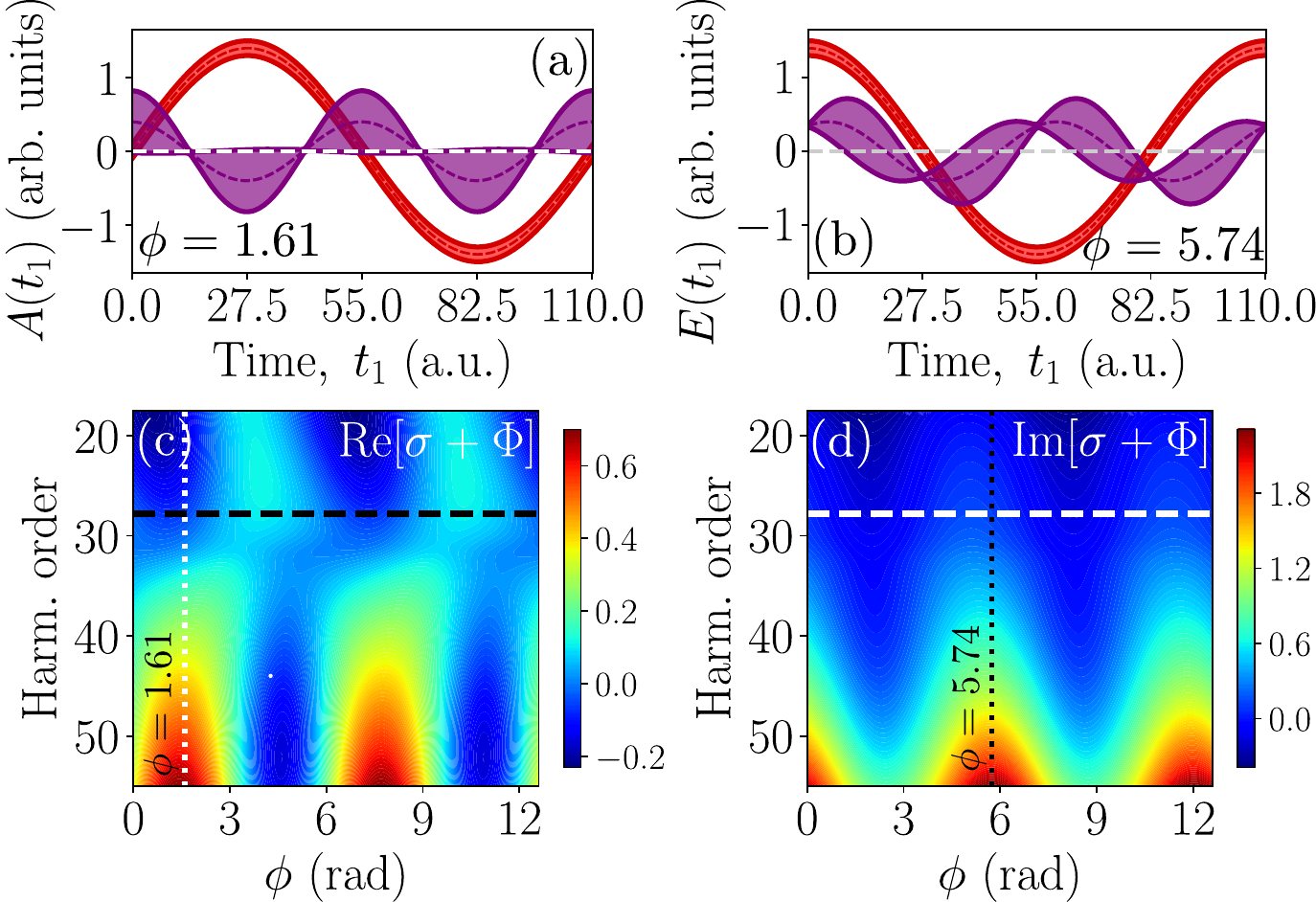}
    \caption{\textbf{Effect of squeezing on the HHG spectrum} Phase dependence of the HHG spectrum when using a squeezed state of light whose squeezing type changes with the two-color delay. Panels (a) and (b) display the vector potential and electric field, respectively, for two different values of $\phi$. Panels (c) and (d) show the real and imaginary part of the dipole phase $\sigma+\Phi$ as a function of $\phi$ when the driving includes such squeezing features. Here, we have set $E_{\omega} = 0.053$ a.u., $E_{2\omega} = 10^{-2} E_\omega$, $\omega = 0.057$ a.u., $I_{\text{squ}} = 10^{-6}$ a.u., and $I_p = 0.9$ a.u., corresponding to Helium atoms.
    }
    \label{Fig:NC:spectrum}
\end{figure}

By evaluating the real and imaginary parts of $\sigma + \Phi$ [Figs.~\ref{Fig:NC:spectrum}~(c) and (d)], we observe that the introduction of squeezing non-trivially modifies the electron trajectories. Specifically, there is a general enhancement (suppression) of the positive (negative) values of both the real and imaginary components. In the case of $\text{Re}[\sigma + \Phi]$, these modifications occur predominantly at values of $\phi$ for which the $2\omega$ field exhibits phase-like squeezing [Fig~\ref{Fig:NC:spectrum}~(a)]. In contrast, for $\text{Im}[\sigma + \Phi]$, the modifications arise in regimes characterized by amplitude-like squeezing [Fig.~\ref{Fig:NC:spectrum}~(b)].

From a physical point of view, the overall increase in $\text{Re}[\sigma + \Phi]$ can be interpreted as a boost in the electron's kinetic energy during the acceleration phase, while the enhancement in $\text{Im}[\sigma + \Phi]$ corresponds to a further suppression of the barrier through which the electron tunnels. For both components, we observe that the maxima are enhanced and the minima are suppressed in regions where the mean value of the field is in-phase (maxima) or out-of-phase (minima) with the strong $\omega$ driver. In the case of maxima enhancement, the effect induced by the squeezed fluctuations tend to constructively interfere with that of the coherent component of the field, effectively increasing the vector potential experienced by the electron at the moment of ionization in the case of $\text{Re}[\sigma + \Phi]$ [Fig.~\ref{Fig:NC:spectrum}~(a)], or lowering the potential barrier in the case of $\text{Im}[\sigma + \Phi]$ [Fig.~\ref{Fig:NC:spectrum}~(b)]. Conversely, for the suppression of the minima---where the mean values of the $\omega$ and $2\omega$ drivers are out-of-phase---the squeezed-fluctuations tend to cancel the coherent contribution of the $2\omega$ field. This results in more efficient electron acceleration and enhanced tunneling probabilities, corresponding to $2\omega$ field configurations with a phase $\pi$ different to those in Fig.~\ref{Fig:NC:spectrum}~(a) and (b).

\section*{Quantum state of the interferometry harmonics}

When both fields are classical coherent states, tuning the phase difference $\phi$ between the $\omega$ and $2\omega$ drivers enables sub-femtosecond control over the HHG dynamics. However, its impact on the quantum state or the photon statistics of the generated harmonics is trivial. 
The post-interacting state remains a product of coherent states, independent of the value of $\phi$. It is therefore natural to ask: \textit{Can the quantum optical attosecond interferometry framework introduced here be leveraged to control the quantum state of the generated harmonics?}

To address this question, we analyze the quantum state of the $q$th harmonic mode as a function of the two-color phase delay. 
The corresponding state is given by 
\begin{equation}
	\hat{\rho}_q(t)
		= \int \dd^2\alpha
			\int  \dd^2 \beta
			\dfrac{P(\alpha,\beta^*)}{\langle\chi^{(\beta^*)}_{q}\vert\chi^{(\alpha)}_{q}\rangle} 
			\relaxket{\chi^{(\alpha)}_{q}}\!\relaxbra{\chi^{(\beta^*)}_{q}},
\end{equation}
where we see that, for sufficiently broad $P(\alpha,\beta^*)$ distributions the phase delay $\phi$ continuously modifies the harmonic amplitudes $\chi^{(\alpha)}_q$. This implies a potential enhancement of field fluctuations in the emitted harmonic radiation, which we asses by evaluating three distinct quantum optical measures:~the Wigner function $W(x_1,x_2)$, the variance along different optical quadratures $\Delta X_i$, and the second-order autocorrelation function $g^{(2)}(0)$. 

Each of these quantities probes different aspects of the quantum optical nature of the harmonics. The Wigner function provides a phase-space representation where negative regions are indicative of non-classical features. The quadrature variances describe the distribution of field fluctuations with $(\Delta X_i)^2 < 0.5$ signaling squeezing below the vacuum fluctuations. Finally, the second-order autocorrelation reveals information about the photon statistics, where values $g^{(2)}(0) < 1$ reveal sub-Poissonian behavior, a hallmark of quantum light. 
These quantities are obtained via
\begin{equation}\label{Eq:QO:observables}
	\langle O_q \rangle
		= \int \dd^2 \alpha
				\int \dd^2 \beta 
					\dfrac{P(\alpha,\beta^*)}{\langle\chi^{(\beta^*)}_{q}\vert\chi^{(\alpha)}_{q}\rangle} 
					f\big(
						\chi_q^{(\beta^*)},
						\chi_q^{(\alpha)}
					 \big),
\end{equation}
where $f(\chi_q^{(\beta^*)},\chi_q^{(\alpha)})$ is a well-defined function of the harmonic amplitudes, whose specific form depends on the quantum optical observable under consideration.

To evaluate this expression, we work under both the classical and thermodynamic limits. The classical limit is appropriate in our context, as we consider fields propagating in the continuum, with the quantization volume tending to infinity $(\kappa \to 0)$, and in regimes where the mean photon number of the driving field $\abs{\alpha}^2$ is extremely large~\cite{gorlach2023high,wang_high-order_2025}. 
In typical HHG experiments the harmonic signal originates from the collective emission of many atoms (usually $10^{10}\sim 10^{12}$), which coherently enhances the dipole response. This can be captured through the rescaling $\chi_q \to N \chi_q$~\cite{lewenstein2021generation,rivera2022strong,stammer2023quantum,stammer2025theory}, where $N$ is the number of atoms in the interaction region.~Therefore, in the thermodynamic limit, we assume that as $\kappa\to 0$ and $N\to \infty$, the atomic density remains constant, i.e., $N\kappa \equiv \varrho = \text{constant}$. Under these conditions, and when considering squeezed drivers, we can rewrite Eq.~\eqref{Eq:QO:observables} in terms of the electric field amplitudes $\varepsilon_{\alpha} = 2 \kappa \alpha$. We denote with $\bar{\varepsilon}_{2\omega,i}$ the mean amplitude of the $2\omega$ driver in the optical quadrature along which anti-squeezing takes place.

\begin{figure}
    \centering
	\includegraphics[width = 0.95\columnwidth]{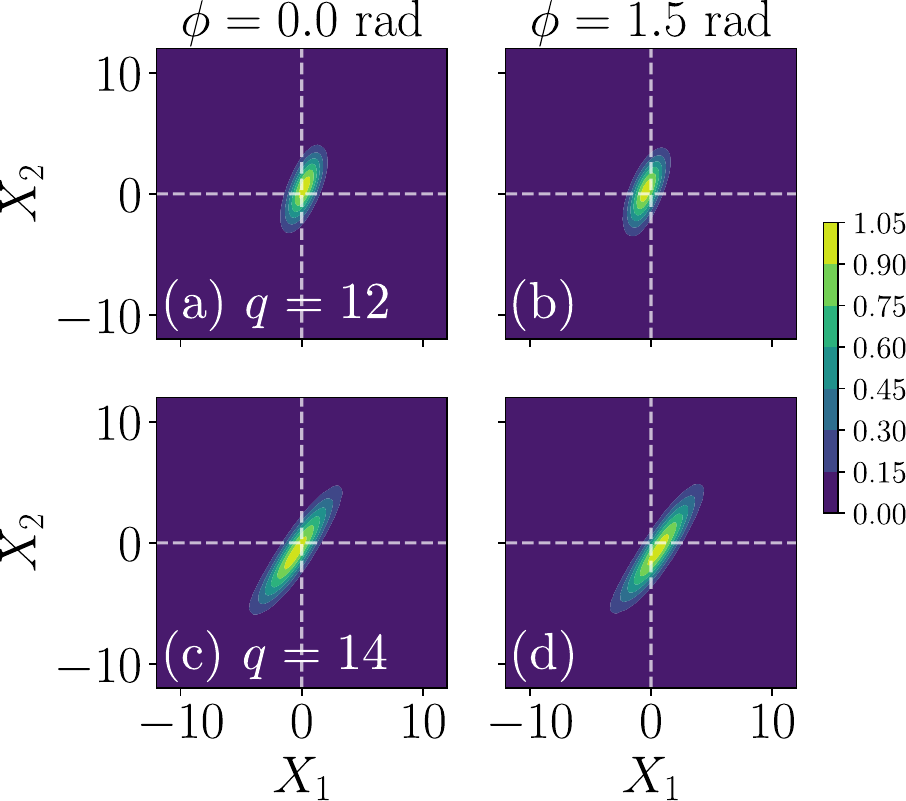}
	\caption{\textbf {Harmonic quantum state} Wigner function of the even harmonics $q=12$ (a)-(b) and $q=14$ (c)-(d) in the attosecond quantum interferometry experiment for two different values of $\phi$, showing non-trivial characteristics by stretching along a field quadrature.~The functions have been normalized to their maximum value.~A more detailed analysis of the quadrature stretching can be seen in Fig.~\ref{fig:g2:and:var}.~The same field parameters as those in Fig.~\ref{Fig:NC:spectrum} were considered here.
    }
      \label{Fig:Harmonics:QO}
\end{figure}

Figure~\ref{Fig:Harmonics:QO} shows the Wigner distribution of different even harmonics for different two-color phase delays $\phi$. In all cases, the Wigner functions exhibit squeezed-like features. These characteristics are notably absent features for odd harmonic orders (see SM), which instead display symmetric, Gaussian-like distributions. This is indeed expected: in the absence of the $2\omega$ driver, the quantum state of the harmonics is given by Eq.~\eqref{Eq:QO:state:coh}, with even orders residing in vacuum states. As such, the small perturbation introduced by the two-color driver does not substantially affect either the amplitude nor quantum state of the odd harmonic orders. In contrast, the even harmonic orders---whose generation relies on the presence of the two-color driver---inherit some of its squeezing characteristics.

\begin{figure}
    \centering
    \includegraphics[width=1\columnwidth]{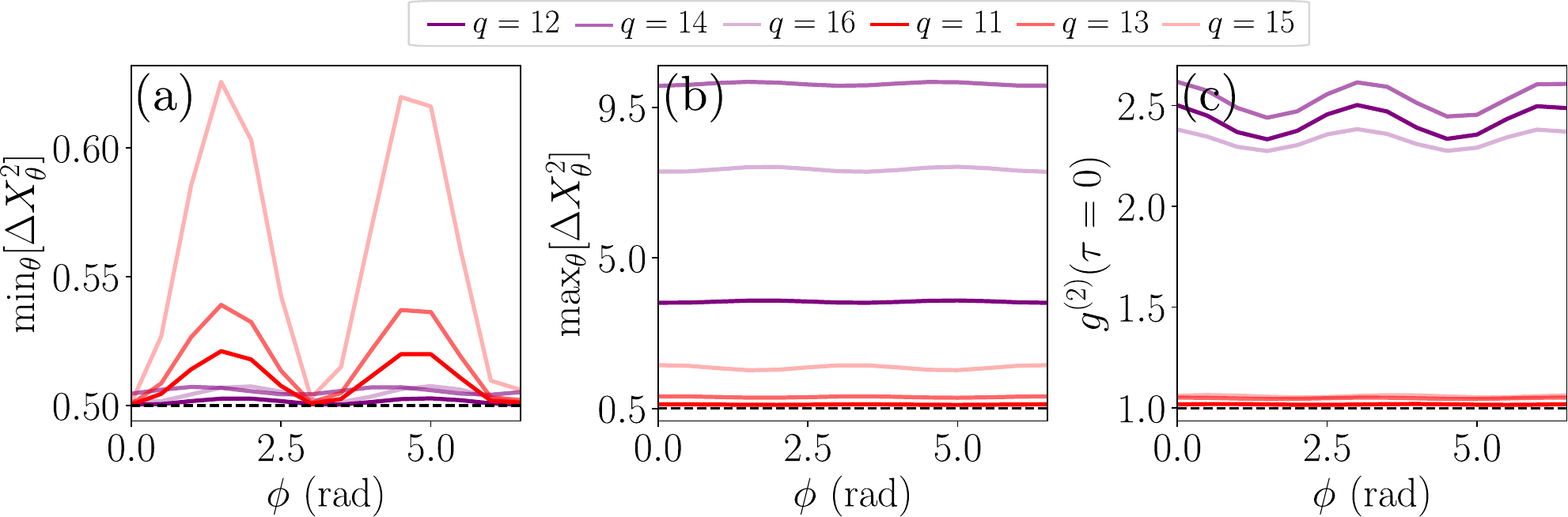}
    \caption{\textbf{Harmonic field observables} Properties of different even (purple colors) and odd (red colors) harmonic orders for varying driving field delay $\phi$. Minimal (a) and maximal (b) variance of the field quadrature $\Delta X_\theta^2$ optimized over the quadrature angle $\theta$. In (c) the second order intensity correlation function $g^{(2)}(\tau = 0)$ showing super-bunching signatures for the even harmonics.}
    \label{fig:g2:and:var}
\end{figure}

Although the Wigner functions of the even orders clearly display features reminiscent of squeezing, they should satisfy the conditions $\max_{\theta}[(\Delta X_{\theta})^2] > 0.5$ and $\min_{\theta}[(\Delta X_{\theta})^2] < 0.5$, where $X_{\theta} = (\hat{a} e^{-i\theta} + \hat{a}^\dagger e^{i\theta})/\sqrt{2}$, such that their product would saturate Heisenberg's uncertainty principle. Figure~\ref{fig:g2:and:var} displays these two quantities for both even and odd harmonic orders, shown in red and purple, respectively. For all harmonics, the minimum variance remains above the vacuum fluctuations of $\min_{\theta}[(\Delta X_{\theta})^2] = 0.5$.
However, for even harmonics the maximum variance is significantly increased. This indicates that the squeezed-like features are not a result of \emph{quantum} squeezing but instead stem from an asymmetric distribution of the field fluctuations across optical quadratures. Interestingly, this noise distribution can be tuned by varying $\phi$, thereby controlling the photon statistics of the generated harmonics. This is illustrated in panel (c), which shows $g^{(2)}(0)$ as a function of $\phi$. The even harmonic orders exhibit a clear super-Poissonian behavior that depends on $\phi$, whereas the odd harmonics remain close to Poissonian, indicative of their coherent nature.

\section{Discussion}

Our work establishes a connection between weak measurement scenarios from quantum measurement theory and the regime of strongly driven systems, demonstrating that this enables the observation of previously inaccessible spectral features.
This is achieved by building a link between weak measurements and attosecond interferometry experiments. 
We have explicitly introduced corrections to the traditional saddle point equations in strong field physics (see Eqs.~\eqref{eq:saddle_point_equations_ionization}-~\eqref{eq:saddle_point_equations_recombination}), revealing the appearance of weak values within strong field physics. 
These corrections allow to obtain spectral information of the electron, such as Fano resonances, which remain unnoticed otherwise. 
We anticipate that the study of different field geometries~\cite{rivera2025structured}, or the presence of spectral features such as field induced resonances~\cite{stammer2020evidence} allow for further insights into these weak measurement corrections.

Extending this scheme towards non-classical driving fields allowed to realize high-photon number quantum states of light with controllable features over a wide spectral range. The corresponding Wigner functions show quadrature stretching differing from those of classical coherent states, while the photon statistics show super-bunching signatures as $g^{(2)}(0) > 2$ across a wide range of harmonic orders.
We dub this approach for generating controllable quantum states of light using strong field physics as \textit{attosecond quantum interferometry (AQI)}.

The use of attosecond quantum interferometry opens the path for several potential applications, ranging from the generation of non-trivial quantum states of light towards the XUV regime or measuring spectral properties hidden from previous approaches.
We furthermore believe, that many attosecond experiments measuring time delays might be interpreted in terms of weak measurements~\cite{maier2025search}. Indeed, such experiments always rely on interferometric set-ups which provide direct access to the phase and therefore to the time~\cite{steinberg1995much}.

Finally, this work suggests that the notion of weak measurement is a powerful technique in order to describe interferometric strong field measurements, since it is ubiquitous in strong field and attosecond physics, and suggests that the notion of weak measurement should be explored in more detail in future studies even beyond the strong field and attosecond physics domain.

The formalism of weak measurement in strong field physics developed in this work can be adopted to a wide range of methods used to analyze strong field phenomena, such as RABBITT~\cite{paul2001observation, muller2002reconstruction}, Attosecond Streaking~\cite{hentschel2001attosecond, constant1997methods, itatani2002attosecond} or KRAKEN~\cite{laurell2025measuring, laurell2022continuous}, all relying on interference effects. With this, the notion of\textit{attosecond quantum interferometry} can provide a new dimension in measuring processes on the attosecond time-scale by using the influence of the ultrafast electron dynamics on the quantum state of the light field.

\ \\

\noindent\textbf{{Acknowledgments}}  
P.S. acknowledges fruitful discussions with Olga Smirnova during his Master Thesis inspiring this work. P. Stammer acknowledges funding from: The European Union’s Horizon 2020 research and innovation programme under the Marie Skłodowska-Curie grant agreement No 847517.
ICFO-QOT group acknowledges support from:
European Research Council AdG NOQIA; MCIN/AEI (PGC2018-0910.13039/501100011033, CEX2019-000910-S/10.13039/501100011033, Plan National FIDEUA PID2019-106901GB-I00, Plan National STAMEENA PID2022-139099NB, I00, project funded by MCIN/AEI/10.13039/501100011033 and by the ``European Union NextGenerationEU/PRTR'' (PRTR-C17.I1), FPI); QUANTERA DYNAMITE PCI2022-132919, QuantERA II Programme co-funded by European Union’s Horizon 2020 program under Grant Agreement No 101017733; Ministry for Digital Transformation and of Civil Service of the Spanish Government through the QUANTUM ENIA project call - Quantum Spain project, and by the European Union through the Recovery, Transformation and Resilience Plan - NextGenerationEU within the framework of the Digital Spain 2026 Agenda; Fundació Cellex; Fundació Mir-Puig; Generalitat de Catalunya (European Social Fund FEDER and CERCA program; Barcelona Supercomputing Center MareNostrum (FI-2023-3-0024); Funded by the European Union. Views and opinions expressed are however those of the author(s) only and do not necessarily reflect those of the European Union, European Commission, European Climate, Infrastructure and Environment Executive Agency (CINEA), or any other granting authority.  Neither the European Union nor any granting authority can be held responsible for them (HORIZON-CL4-2022-QUANTUM-02-SGA  PASQuanS2.1, 101113690, EU Horizon 2020 FET-OPEN OPTOlogic, Grant No 899794, QU-ATTO, 101168628),  EU Horizon Europe Program (This project has received funding from the European Union’s Horizon Europe research and innovation program under grant agreement No 101080086 NeQSTGrant Agreement 101080086 — NeQST); ICFO Internal ``QuantumGaudi'' project; 
M. F. C. acknowledges financial support from the National Key Research and Development Program of China (Grant No.~2023YFA1407100), Guangdong Province Science and Technology Major Project (Future functional materials under extreme conditions - 2021B0301030005) and the Guangdong Natural Science Foundation (General Program project No. 2023A1515010871).
\\

\noindent\textbf{{Author contributions}}
P.S. and J.R.-D. contributed equally to this work.
P.S.: Conceived the idea, developed the theoretical approach and supervised this work. J.R.-D.: Performed the numerical simulation of this work and contributed to the theoretical calculations. 
P.S. and J.R.-D. wrote the manuscript. 
M.L. contributed to parts of the idea of this work.
Part I of the Results section was adapted from P.S. Master Thesis.\\

\ \\
\noindent\textit{{Note added:}}  
During the final stages of completing this manuscript, a paper with a similar field configuration appeared on the arXiv [arXiv:2502.09427]. \\

\bibliography{literature}{}

\appendix

\section*{Appendix}

\subsection*{Derivation of perturbed dipole moment}

We consider the perturbed dipole moment expectation value $\expval{\vb{d}'(t)} = \bra{\psi'(t)} \vb{d} \ket{\psi'(t)}$, where $\ket{\psi'(t)} = U_{sc}' (t) \ket{g}$. We will use that the evolution of the electronic wavefunction $\ket{\psi'(t)}$ under the full semi-classical Hamiltonian (including the perturbation) is known \cite{lewenstein1994theory}, such that we can write the generic solution of the full dynamics as 
\begin{align}
    \ket{\psi^\prime(t)} = & U_0(t,t_0) \ket{g} \nonumber \\
    & - i \int_{t_0}^t dt_1 U^\prime_{sc}(t,t_1) V_L^\prime(t_1) U_0(t_1,t_0) \ket{g},
\end{align}
where $U_0(t,t_0)$ is the propagator under the atomic Hamiltonian $H_a$ alone, and
\begin{align}
    V_L^\prime(t) = - \vb{d} \cdot [\vb{E}_\omega(t) + \vb{E}_{2\omega}(t)],
\end{align}
is the total semi-classical interaction with the classical field
\begin{align}
        \vb{E}_{cl}(t) & = \vb{E}_\omega(t) + \vb{E}_{2\omega}(t) \\
        & =  2 \kappa f(t) \left[ \abs{\alpha_1} \sin(\omega t) + \abs{\alpha_2} \sin(2\omega t + \phi) \right]. \nonumber
\end{align}

To obtain the perturbed dipole moment we employ the commonly done in strong laser field physics Strong Field Approximation (SFA) \cite{lewenstein1994theory, amini2019symphony} by neglecting the attractive Coulomb potential of the core after ionization such that the semi-classical propagator can be written as a Volkov propagator, i.e. $U_{sc}^\prime(t,t_1) \simeq U_V^\prime(t,t_1)$. Furthermore, we approximate the exact continuum state with momentum $\vb{k}$ with plane wave Volkov states of kinetic momentum $\vb{k} = \vb{p}+ \vb{A}^\prime(t)$, where $\vb{p}$ is the canonical momentum and $\vb{A}^\prime(t) = \vb{A}_\omega(t) + \vb{A}_{2\omega}(t)$ is the total vector potential including the second harmonic perturbation.

We shall now look at the perturbation to the transition dipole moment by expanding the perturbed Volkov state $\ket{\vb{p} + \vb{A}'(t)}$, which includes the total vector potential, up to first order 
\begin{align}
    \ket{\vb{p} + \vb{A}^\prime(t) } & \simeq  \ket{\vb{p} + \vb{A}(t)}  + \epsilon \partial_\epsilon \ket{\vb{p} + \vb{A}^\prime(t) } |_{\epsilon =0} \\
    & =  [ 1+ i \vb{A}_{2\omega}(t)  \cdot \vb{d}] \ket{\vb{p} + \vb{A}(t)} + \mathcal{O}(\epsilon^2), \nonumber
\end{align}
such that we can write (in terms of components $i \in \{ x,y,z \}$)
\begin{align}
    \bra{g} d_i \ket{\vb{p} + \vb{A}^\prime(t)} \simeq \bra{g} d_i  \ket{\vb{p} + \vb{A}(t)} e^{i {A}_{2\omega}(t)   D_{ij}(t)},
\end{align}
where we have defined the weak value for the recombination transition matrix element 
\begin{align}
\label{eq:weak_value_general_app}
    D_{ij}(t) =   \frac{\bra{g}   d_i d_j   \ket{\vb{p} + \vb{A}(t)}  }{\bra{g} d_i \ket{\vb{p} + \vb{A}(t)}}. 
\end{align}
where $d_j = \boldsymbol{\epsilon}_2 \cdot \vb{d} $ with $\boldsymbol{\epsilon}_{2}$ the polarization unit vector of the $2\omega$ field.
This is the weak value of the dipole moment for the recombination transition matrix element. The WV is in general complex valued~\cite{dressel2014colloquium, dressel2015weak}, and here, a vectorial quantity along the polarization direction of the $2\omega$ field.
We now can proceed along the same lines for the transition matrix element of ionization which is given by 
\begin{align}
    & \bra{\vb{p} + \vb{A}'(t_1)}   V^\prime_L(t_1) \ket{g}  \simeq  \bra{\vb{p} + \vb{A}(t_1)} V_L(t_1) \ket{g} \\
    & \quad \times \exp[- i A_{2\omega}(t_1) D_{ji}(t_1)] \exp[ \Delta F(t_1, \phi)], \nonumber
\end{align}
where the ionization WV satisfies $D_{ji}(t) = D_{ij}^*(t)$, and 
\begin{align}
    \Delta F (t_1, \phi) = \frac{F_{2\omega}(t_1, \phi)}{F_\omega(t_1)} = \epsilon \frac{\sin(2 \omega t_1 + \phi)}{\sin(\omega t_1)}.
\end{align}

The second factor only appears in the ionization matrix element which involves the tunneling process, which leads to a change in the tunnel ionization probability due to the second perturbative field.

\subsection*{New phase correction terms}

Within the SFA~\cite{amini2019symphony}, the dipole moment is given by
\begin{align}
\label{eq:dipole_SFA_perturbed}
    \expval{\vb{d}'(t)} = & - i \int_{t_0}^t dt_1 \int d\vb{p} \bra{g} \vb{d} \ket{\vb{p} + \vb{A}(t)} e^{- i \mathcal{S}(t,t_1,\vb{p}, \phi)} \nonumber \\
    & \times \bra{\vb{p} + \vb{A}(t_1)} V_L(t_1) \ket{g}  e^{\Delta F(t_1, \phi) +\operatorname{Im}\Phi} + c.c.,
\end{align}
where the total phase reads 
\begin{align}
\label{eq:phase_total}
    \mathcal{S} ( t,t_1, \vb{p}, \phi) = & S( t, t_1,\vb{p}) + \Delta S( t, t_1,\vb{p}, \phi).
\end{align}
Here, $S( t, t_1,\vb{p}) = S_0( t, t_1,\vb{p}) + I_p (t-t_1)$ is the usual phase considered in the process of HHG \cite{lewenstein1994theory}, and the correction term is given by 
\begin{align}
    \Delta S( t, t_1,\vb{p}, \phi) = \sigma( t,t_1, \vb{p}, \phi) + \operatorname{Re}[\Phi( t,t_1, \vb{p}, \phi)].
\end{align}

The correction to the phase due to the $2\omega$ perturbation comes from an energy correction in the action during the propagation in the continuum $\sigma( t,t_1, \vb{p}, \phi)$, and a second phase term $\Phi ( t,t_1, \vb{p}, \phi)$.
One of the main new results in the WM scenario considered here, is an additional phase contribution from the ionization and recombination matrix elements
\begin{align}
\label{eq:correation_phase2}
    \Phi ( t,t_1, \vb{p}, \phi) = A_{2\omega}(t_1, \phi) D_{ji}(t_1) - A_{2\omega} (t, \phi) D_{ij}(t).
\end{align}

\subsection*{Quantum state under non-classical drive}

The evolved joint light-matter system can be written as
\begin{equation}\label{Eq:Total:state:light:matter}
	\begin{aligned}
		\hat{\rho}(t)
			= \int \dd^2\alpha\int \dd^2 \beta
				\dfrac{P(\alpha,\beta^*)}{\braket{\beta^*}{\alpha}}
					\Big[&
						\dyad{\phi_{\alpha}(t)}{\phi_{\beta^*}(t)}
						\\&\otimes \dyad{\Phi_{\alpha}(t)}{\Phi_{\beta^*}(t)}
						\Big],
	\end{aligned}
\end{equation}
where $\ket{\phi_{\alpha}(t)}$ denotes the quantum state of the electron, obtained through standard semiclassical HHG analyses~\cite{lewenstein1994theory,amini2019symphony}.~The state $\ket{\Phi_{\alpha}(t)}$ is formally analogous to that in Eq.~\eqref{Eq:QO:state:coh}, with the key distinction that the time-dependent dipole $\langle \vb{d}(t)\rangle$ leading to Eq.~\eqref{eq:dipole_SFA_perturbed} is now evaluated under the classical coherent field $\vb{E}^{(\alpha)}_{\text{cl}}(t) = \Tr[\vb{E}_Q(t) \dyad{\alpha_1}\otimes \dyad{\alpha} \dyad{\{0_q\}}]$.

\subsection*{Evaluation of quantum optical observables}

Following Ref.~\cite{tzur2024generation}, in the low-depletion regime, the final quantum optical state associated to the $q$th harmonic mode, after the interaction with an ensemble of $N$ atoms~\cite{stammer2023quantum}, can be expressed as
\begin{equation}\label{Eq:Meth:state}
    \begin{aligned}
	\hat{\rho}_q(t)
		= \int \dd^2\alpha \int \dd^2 \beta& \dfrac{P(\alpha,\beta^*)}{\braket{N\chi_{\beta^*,q}(t)}{N\chi_{\alpha,q}(t)}}
        \\&\times
			\dyad{N\chi_{\alpha,q}}{N\chi_{\beta^*,q}},
    \end{aligned}
\end{equation}
from which any quantum optical observable $\hat{O}$ acting on the harmonic mode $q$ can be computed as
\begin{equation}\label{Eq:Meth:Obs}
    \begin{aligned}
	\langle\hat{O}\rangle_q
		= \int \dd^2\alpha \int \dd^2 \beta&
				\dfrac{P(\alpha,\beta^*)}{\braket{N\chi_{\beta^*,q}(t)}{N\chi_{\alpha,q}(t)}}
				\\&\times 	o(N\chi_{\alpha,q},N\chi_{\beta^*,q}),
    \end{aligned}
\end{equation} 
where we assume that the observable $\hat{O}$ does not introduce any additional dependencies on either the number of emitters $N$ nor the light-matter coupling parameter $\kappa$.

To evaluate these quantum optical observables, we work both in the classical and thermodynamic limits, defined as follows:
\begin{itemize}
	\item \textbf{Classical limit.} In this regime, we express the coherent state amplitude $\alpha = 2\kappa \varepsilon_{\alpha}$, where $\varepsilon_\alpha$ denotes the electric field amplitude. This limit entails setting $V\to \infty$ and $\kappa \to \infty$, the first motivated by the fact that we are dealing with fields propagating in free space, where the quantization volume $V\to\infty$ (implying $\kappa\to0$). Consequently, to maintain a finite electric field amplitude, one must take $\alpha \to \infty$.
	\item \textbf{Thermodynamic limit.} Since $V\to \infty$, a non-vanishing harmonic generation signal requires $N\to \infty$, such that the atomic density in the interaction region $\varrho = \epsilon N $ remains finite. As a result, the coherent state amplitude associated with the harmonic mode is given by $\chi_q = \sqrt{q} N\kappa \langle d(\omega_q)\rangle = \sqrt{q}\varrho\langle d(\omega_q)\rangle\equiv \varrho_q\langle d(\omega_q)\rangle$.
\end{itemize}

Under these conditions, Eq.~\eqref{Eq:Meth:Obs} becomes
\begin{equation}
	\begin{aligned}
	\langle \hat{O}_q\rangle
		&= 
			\int \dd^2 \varepsilon_\alpha
				\int \dd^2 \varepsilon_\beta
					\bigg[
						\lim_{\kappa\to0} \dfrac{1}{16\kappa^4}P(\varepsilon_\alpha,\varepsilon^*_\beta)
					\bigg]
					\\&\hspace{2cm}\times
					\dfrac{o(\varrho_q\langle d_\alpha(\omega_q)\rangle,\varrho_q\langle d_{\beta^*}(\omega_q)\rangle)}{\braket{\varrho_q\langle d_{\beta^*}(\omega_q)\rangle}{\varrho_q\langle d_\alpha(\omega_q)\rangle}},
	\end{aligned}
\end{equation}
where, in the case of using squeezed light, the limiting behavior of the distribution $P(\alpha,\beta^*)$ is given by~\cite{even_tzur_photon-statistics_2023,rivera2025structured}
\begin{equation}
	\begin{aligned}
	\lim_{\kappa\to0} \dfrac{1}{16\kappa^4}P(\varepsilon_\alpha,\varepsilon^*_\beta)
		&= \dfrac{1}{\sqrt{2\pi \varsigma_i}}
				\exp[-\dfrac{\big(\varepsilon_{\alpha,i}-\bar{\varepsilon}_{i}\big)}{2\varsigma_i}]
				\\&\quad\times
				\delta(\varepsilon_\alpha - \varepsilon_\beta^*)
				\delta(\varepsilon_{\alpha,\bar{i}} - \bar{\varepsilon}_{\bar{i}}).
	\end{aligned}
\end{equation}
Here, $\bar{i}$ denotes the phase-space direction along which the squeezing is applied, $i$ the orthogonal direction, $\bar{\varepsilon}_{i}$ and $\bar{\varepsilon}_{\bar{i}}$ the coherent state amplitudes along each respective axis, and $\varsigma_i = 4 I_{\text{squ}}$. quantifies the increased field fluctuations along direction $i$.

It is important to emphasize at this point that the classical and thermodynamic limits discussed above apply strictly to the harmonic modes. If the same limits are applied directly to the externally prescribed driving field mode, significant limitations emerge which---if not treated carefully---can lead to incorrect properties results, as recently pointed out in Ref.~\cite{gothelf2025high}. This issue arises because quantum optical observables, such as the photon number operator, are expressed in terms of creation and annihilation operators acting on the driving field mode, whose state is expressed in our case as
\begin{equation}
	\hat{\rho} 
		= \int \dd^2\alpha \int \dd^2 \beta
			 \dfrac{P(\alpha,\beta^*)}{\braket{\beta^*}{\alpha}}
			 	\dyad{\alpha}{\beta^*}.
\end{equation}
For example, the mean photon number takes the form
\begin{equation}
	\langle \hat{a}^\dagger \hat{a}\rangle
		= \int \dd^2 \alpha \int \dd^2 \beta P(\alpha,\beta^*)\alpha \beta, 
\end{equation}
which is a well-defined quantity and leads to the expected results for the considered initial state. However, if one aims to apply the classical limit, now acting on both the $P(\alpha,\beta^*)$ and $\alpha\beta$ contributions, the resulting expression can diverge. Unlike the harmonic modes, such divergences cannot be regularized via the thermodynamic limit, since the driving field is not a collective quantity arising from many-body contributions. Instead, it is a prescribed input to the system, and thus lacks the extensive scaling with $N$ that stabilizes observables associated with the emitted harmonics.

\newpage
\onecolumngrid
\setcounter{section}{0}
\begin{center}
        \textbf{SUPPLEMENTARY MATERIAL}
\end{center}

\tableofcontents

\section{Analytical calculation for $D_{ii}(t)$}
A central quantity in our analysis is $D_{ij}(t)$, which provides a modification to the standard semiclassical action when adding the $2\omega$ field on top of the $\omega$ component. From a more practical point of view, this terms arises when evaluating the matrix elements of the dipole moment operator between the ground state and the exact continuum states $\ket{\vb{p} + \vb{A}(t)}$. Since the contribution of $\vb{A}_{2\omega}(t)$ is perturbative compared to the $\vb{A}_{\omega}(t)$ one, we can approximately write
\begin{equation}
	\ket{\vb{p} + \vb{A}(t)} 
		\simeq 
			[1 + i \vb{A}_{2\omega}(t) \cdot \vb{d}]
				\ket{\vb{p} + \vb{A}_{\omega}(t)}
		= [1 + i A_{2\omega}(t) d_j]
				\ket{\vb{p} + \vb{A}_{\omega}(t)},
\end{equation}
where we have assumed that the $2\omega$ component is linearly polarized along the $j$-direction. Therefore, we can write the transition matrix element for the $\hat{d}_i$ operator as
\begin{align}
	\mel{g}{\hat{d}_i}{\vb{p}+\vb{A}(t)}
		&= \mel{g}{\hat{d}_i}{\vb{p} + \vb{A}_\omega(t)}
			+ i A_{2\omega}(t) \mel{g}{\hat{d}_i\hat{d}_j}{\vb{p} + \vb{A}_\omega(t)}
		\\&=  \mel{g}{\hat{d}_i}{\vb{p} + \vb{A}_\omega(t)}
			\big[ 1 + i A_{2\omega}(t) D_{ij}(t) \big],\label{Eq:dipole:second}
\end{align}
where $D_{ij}(t)$ is defined as in the main text, that is, as
\begin{equation}\label{Eq:def:Dij}
	D_{ij}(t)
		= \dfrac{\mel{g}{\hat{d}_i\hat{d}_j}{\vb{p} + \vb{A}_\omega(t)}}{\mel{g}{\hat{d}_i}{\vb{p} + \vb{A}_\omega(t)}}.
\end{equation}

For standard strong-field conditions---$E_\omega = 0.053$ a.u. and $\omega= 0.057$ a.u.---the amplitude of the vector potential for the $\omega$-field component is $A_{0,\omega} \approx 1$ a.u., and in our case $A_{0,2\omega} = \epsilon A_{0,\omega}$ with $\epsilon \ll 1$.~Therefore, we can try to approximate
\begin{equation}
    1 + i A_{2\omega}(t) D_{ij}(t) 
        \approx e^{iA_{2\omega}(t) D_{ij}(t)},
\end{equation}
which is to good extent valid when $\abs{A_{2\omega}(t) D_{ij}(t)} \ll 1$. Since $A_{2\omega}(t)$ is a perturbative quantity, the general condition for this approximation to hold is
\begin{equation}
	\mel{g}{\hat{d}_i}{\vb{p} + \vb{A}_\omega(t)}
		\gg A_{2\omega(t)} \mel{g}{\hat{d}_i\hat{d}_j}{\vb{p} + \vb{A}_\omega(t)},
\end{equation}
which, however, does not hold at points where $\mel{g}{\hat{d}_i}{\vb{p} + \vb{A}_\omega(t)} = 0$. This turns crucial when doing saddle-point analyses, as when deforming our integration contour the contributions of poles in $D_{ij}(t)$ might give rise to regions where derivatives of modified versions of the action become arbitrarily small, though do not correspond to actual saddle-points.

Hereupon, we assume that the two fields are linearly polarized along the same direction, meaning that $i=j$, such that Eq.~\eqref{Eq:def:Dij} reads
\begin{equation}
	D_{ii}(t) = \dfrac{\langle g\rvert \hat{d}_i^2 \lvert p +A_{\omega}(t) \rangle}{\langle g\rvert \hat{d}_i \lvert p +A_{\omega}(t) \rangle},
\end{equation}
where $A_{\omega}(t)$ is the vector potential corresponding to the $\omega$-field component. Given that in general we can find analytical functions of $\langle g\vert \hat{d}_i \vert p+A_{\omega}(t)\rangle$, let us rewrite the numerator above such that these contributions are explicitly included. To do so, we introduce the SFA version of the identity $\mathbbm{1} = \dyad{g} + \dyad{p + A_{\omega}(t)}$, allowing us to express
\begin{align}
	\langle g\rvert \hat{d}_i^2 \lvert p +A_{\omega}(t) \rangle
		&= \langle g\rvert \hat{d}_i \lvert g\rangle \langle g\rvert \hat{d}_i \lvert p+A_{\omega}(t)\rangle
		+ \int \dd p'
				\langle g\rvert \hat{d}_i \lvert p' + A_{\omega}(t)\rangle \langle p' + A_{\omega}(t)\rvert \hat{d}_i \lvert p+A_{\omega}(t)\rangle
		\\&
		= \int \dd p'
		\langle g\rvert \hat{d}_i \lvert p' + A_{\omega}(t)\rangle \langle p' + A_{\omega}(t)\rvert \hat{d}_i \lvert p+A_{\omega}(t)\rangle,
\end{align}
where in transitioning from the first to the second equality we assume that the ground state has a well-defined parity, implying $\langle g\rvert \hat{d}_i \lvert g\rangle = 0$. For the continuum-continuum transition matrix element, we take into account that $\mel{p}{\hat{d}_i}{p'} = i \pdv{}{p_i} \delta(p-p')$, and that $\mel{p}{\hat{d}_i}{p'}^* = \mel{p'}{\hat{d}_i}{p}$, such that we can write
\begin{align}
\label{Eq:mel:d2}
	\langle g\rvert \hat{d}_i^2 \lvert p +A_{\omega}(t) \rangle
		&= \int \dd p' 
				\langle g\rvert \hat{d}_i \lvert p' + A_{\omega}(t)\rangle
				\bigg[
					i \pdv{}{p} \delta(p-p')
				\bigg]^*
		= -i \pdv{}{p}\int \dd p' 
					\langle g\rvert \hat{d}_i \lvert p' + A_{\omega}(t)\rangle
					\delta(p-p')
		\\&= -i \pdv{}{p}
				\langle g\rvert \hat{d}_i \lvert p + A_{\omega}(t)\rangle,
\end{align}
which provides a straightforward way of computing the matrix element of interest. In the following, we evaluate the two cases of interest here: with and without the Fano resonance.

\subsection{Expressions without the Fano resonance}
Following Ref.~\cite{lewenstein1994theory}, we model our atomic potential using a truncated harmonic-oscillator potential, such that this transition matrix element takes the form
\begin{equation}\label{Eq:dip}
	\langle g\rvert \hat{d}_i \lvert p\rangle
		= i \bigg( \dfrac{1}{\pi \alpha}\bigg)^{3/4}
			\dfrac{p}{\alpha}
			\exp[-\dfrac{p^2}{2\alpha}],
\end{equation}
with $\alpha = 0.8 I_p$, showing a Gaussian profile. While any potential, including Coulomb potentials, could be chosen, we adopt this one as saddle-points obtained using standard Strong-Field Approximation~(SFA) do not match well with Coulomb potentials. Addressing Coulomb potentials would require alternative approaches, such as the CQSFA, although soft-core potentials could also be employed

In this case our transition matrix element reads
\begin{equation}\label{Eq:dip2}
	\langle g\rvert \hat{d}_i^2 \lvert p +A_{\omega}(t) \rangle
		=  \bigg( \dfrac{1}{\pi \alpha}\bigg)^{3/4}
			\bigg(
				\dfrac{1}{\alpha}
				- \dfrac{\big(p+A_{\omega}(t)\big)^2}{\alpha^2}
			\bigg)
			\exp[-\dfrac{\big(p+A_\omega(t)\big)^2}{2\alpha}],
\end{equation}
and we therefore find
\begin{equation}\label{Eq:Dii:def}
	D_{ii}(t)
		= \dfrac{-i}{p+A_{\omega}(t)}
			\bigg[
				1 - \dfrac{1}{\alpha}\big(p+A_{\omega}(t)\big)^2
			\bigg].
\end{equation}

\subsection{Expressions with the Fano resonance}
Following Ref.~\cite{rzazewski1983photoexcitation,lewenstein1983photon}, we generalize the dipole moment matrix elements in the presence of a Fano resonance as
\begin{equation}\label{Eq:Fano:dip}
	d_{\text{F}}(v)
		= \dfrac{d(v)}{\sqrt{4\pi \Gamma}}
			\dfrac{\Gamma}{v^2/2 - \omega - i \Gamma}
			- \dfrac{i}{1-iq},
\end{equation}
where $\Gamma$ is related to the lifetime of the autoionizing state and $q$ is the so-called Fano asymmetry parameter.~When comparing Eq.~\eqref{Eq:Fano:dip} with the more general expressions presented in Refs.~\cite{rzazewski1983photoexcitation,lewenstein1983photon}, in the expression above we have already taken $\Gamma_1\to \infty$.~

However, we are not interested only on the matrix elements of $\hat{d}$, but also of $\hat{d}^2$. Making use of Eq.~\eqref{Eq:mel:d2}, we can represent these in terms of derivatives of the matrix elements of $\hat{d}$, which for the case of the Fano resonance results in
\begin{equation}
	\begin{aligned}
		\mel{g}{\hat{d}^2}{p + A_{\omega}(t)}_{\text{F}}
			= \dfrac{\mel{g}{\hat{d}^2}{p + A_{\omega}(t)}}{\sqrt{4\pi \Gamma}}
			\dfrac{\Gamma}{[p+A_{\omega}(t)]^2/2 - \omega - i \Gamma}
			+ i \dfrac{d(v)}{\sqrt{4\pi \Gamma}}
			\dfrac{2[p+A_{\omega}(t)]\Gamma}{\{[p+A_{\omega}(t)]^2/2 - \omega - i \Gamma\}^2},
	\end{aligned}
\end{equation}
where $d(v)$ and $\mel{g}{\hat{d}^2}{p + A_{\omega}(t)}$ are respectively given in Eqs.~\eqref{Eq:dip} and \eqref{Eq:dip2}.~Consequently, the Fano-modified version of $D_{ii}(t)$ reads
\begin{equation}
	D_{ii}^{(\text{F})}(t)
		= \dfrac{\mel{g}{\hat{d}^2}{p + A_{\omega}(t)}_{\text{F}}}{d_{\text{F}}(t)}.
\end{equation} 

\section{Saddle-point analysis}
Our main focus here is to analyze the saddle-point solutions in the absence of the Fano resonance. We then justify why a similar analysis cannot be done in the presence of the Fano resonance, and how we circumvent this issue when computing the spectrum.

\subsection{In the absence of the Fano resonance}
Following the main text, when incorporating the influence of $D_{ii}(t)$ into the action leads to a modified version given by
\begin{equation}
	\mathcal{S}(p,t,t')
	= S_0(p,t,t') + I_p(t-t') + \sigma(p,t,t',\phi) + \Phi(p,t,t',\phi),
\end{equation}
where the phase factor $\Phi (p,t,t',\phi)$ is a perturbative term that modifies the electronic trajectories, given by
\begin{equation}
	\Phi (p,t,t',\phi)
	= A_{2\omega}(t',\phi) D_{ii}(t') - A_{2\omega}(t,\phi) D_{ii}(t).
\end{equation}

To compute the saddle points of this expression, we need to evaluate the derivative of $\Phi(p, t, t', \phi)$ with respect to the integration variables. In the following, we perform this calculation using Eq.~\eqref{Eq:Dii:def} and by expressing the vector potential as $A_{2\omega}(t) = -\tfrac{E_{2\omega}}{2\omega}\sin(2\omega t + \phi)$ and $A_{\omega}(t) = -\tfrac{E_{\omega}}{\omega}\sin(\omega t)$.

\subsubsection{Partial derivative with respect to $t$}
Applying the chain rule, we can write express this partial derivative as
\begin{equation}
	\pdv{\Phi}{t}
	= - \pdv{A_{2\omega}(t,\phi)}{t}D_{ii}(t)
	- A_{2\omega}(t,\phi) \pdv{D_{ii}(t)}{t}
	= E_{2\omega}\cos(2\omega t)D_{ii}(t) 
	- A_{2\omega}(t,\phi) \pdv{D_{ii}(t)}{t},
\end{equation}
and for the partial derivative of $D_{ii}(t)$ we find
\begin{align}
	\pdv{D_{ii}(t)}{t}
	&= \dfrac{-iE_{\omega}\cos(\omega t)}{\big(p+A_{\omega}(t)\big)^2}
	\bigg[
	1 - \dfrac{1}{\alpha}\big(p+A_{\omega}(t)\big)^2
	\bigg]
	+\dfrac{-iE_{\omega}\cos(\omega t)}{\alpha}.
	\\&
	= -i E_{\omega}\cos(\omega t)
	\Bigg\{
	\dfrac{1}{\big(p+A_{\omega}(t)\big)^2}
	\bigg[
	1 - \dfrac{1}{\alpha}\big(p+A_{\omega}(t)\big)^2
	\bigg]
	+ \dfrac{2}{\alpha}
	\Bigg\},
\end{align}
such that the complete function reads
\begin{equation}
	\pdv{\Phi}{t}
	= E_{2\omega}\cos(2\omega t)D_{ii}(t) 
	+iA_{2\omega}(t,\phi)
	E_{\omega}\cos(\omega t)
	\Bigg\{
	\dfrac{1}{\big(p+A_{\omega}(t)\big)^2}
	\bigg[
	1 - \dfrac{1}{\alpha}\big(p+A_{\omega}(t)\big)^2
	\bigg]
	+ \dfrac{2}{\alpha}
	\Bigg\}.
\end{equation}

\subsubsection{Partial derivative with respect to $t'$}
The result is very similar since the expression only differs in a minus sign. We thus get
\begin{equation}
	\pdv{\Phi}{t'}
	= -E_{2\omega}\cos(2\omega t')D_{ii}(t') 
	-iA_{2\omega}(t',\phi)
	E_{\omega}\cos(\omega t')
	\Bigg\{
	\dfrac{1}{\big(p+A_{\omega}(t')\big)^2}
	\bigg[
	1 - \dfrac{1}{\alpha}\big(p+A_{\omega}(t')\big)^2
	\bigg]
	+ \dfrac{2}{\alpha}
	\Bigg\}.
\end{equation}

\subsubsection{Partial derivative with respect to $p$}
In this case, the saddle-point equation reads
\begin{equation}
	\pdv{\Phi}{p}
	= A_{2\omega}(t',\phi) \pdv{D_{ii}(t')}{p}
	- A_{2\omega}(t,\phi) \pdv{D_{ii}(t)}{p},
\end{equation}
where we have that
\begin{equation}
	\pdv{D_{ii}(t)}{p}
	= \dfrac{i}{\big(p+A_{\omega}(t)\big)^2}
	\bigg[
	1 - \dfrac{1}{\alpha}\big(p+A_{\omega}(t)\big)^2
	\bigg]
	+ \dfrac{i2}{\alpha},
\end{equation}
and therefore
\begin{equation}
	\begin{aligned}
		\pdv{\Phi}{p}
		&= A_{2\omega}(t',\phi)
		\Bigg\{
		\dfrac{i}{\big(p+A_{\omega}(t')\big)^2}
		\bigg[
		1 - \dfrac{1}{\alpha}\big(p+A_{\omega}(t')\big)^2
		\Bigg]
		+ \dfrac{i2}{\alpha}
		\bigg\}
		\\&\quad
		- A_{2\omega}(t,\phi)
		\Bigg\{
		\dfrac{i}{\big(p+A_{\omega}(t)\big)^2}
		\bigg[
		1 - \dfrac{1}{\alpha}\big(p+A_{\omega}(t)\big)^2
		\bigg]
		+ \dfrac{i2}{\alpha}
		\Bigg\}.
	\end{aligned}
\end{equation}

\subsubsection{Solving the saddle-point equations}
With these equations, we now proceed to solve the saddle-point equations numerically using the tools provided by the \texttt{RB-SFA} Mathematica package~\cite{RBSFA}. In principle, the $D_{ii}(t)$ function (and consequently $\Phi$) we computed is purely imaginary. This implying that, following Eqs.~(8)-(10) of the main text, it does not contribute to the saddle-points. Nevertheless, we can still encapsulate $\Phi$ into the action as an imaginary term and investigate how it perturbs the trajectories

The above means that the equations we are going to solve are slightly different from the ones in Eqs.~(8)-(10) of the main text, and are specifically given by
\begin{align}
	&\dfrac{\big[p_s + A_\omega(t_{\text{ion}})]^2}{2}
		+ I_p
			= - [p_s + A_{\omega}(t_{\text{ion}})] A_{2\omega} (t_{\text{ion}})
				+ \pdv{\Phi}{t'}\Big\lvert_{\boldsymbol{\theta_s}},~\label{Eq:ionization}
	\\&
	\int^{t_{\text{re}}}_{t_{\text{ion}}} \dd \tau
		\big[p_s + A_{\omega}(\tau)]
			= - \int^{t_{\text{re}}}_{t_{\text{ion}}} \dd \tau A_{2\omega}(\tau)
				- \pdv{\Phi}{p}\Big\lvert_{\boldsymbol{\theta_s}},\label{Eq:propagation}
	\\&
		\dfrac{\big[p_s + A_\omega(t_{\text{re}})]^2}{2}
				+ I_p -q \omega
				= - [p_s + A_{\omega}(t_{\text{re}})] A_{2\omega} (t_{\text{re}})
				- \pdv{\Phi}{t}\Big\lvert_{\boldsymbol{\theta_s}},\label{Eq:recombination}
\end{align}
where $\boldsymbol{\theta_s} = (p_s, t_{\text{re}},t_{\text{ion}})$ denote the saddle-points.

\begin{figure}
	\centering
	\includegraphics[width = 0.55\textwidth]{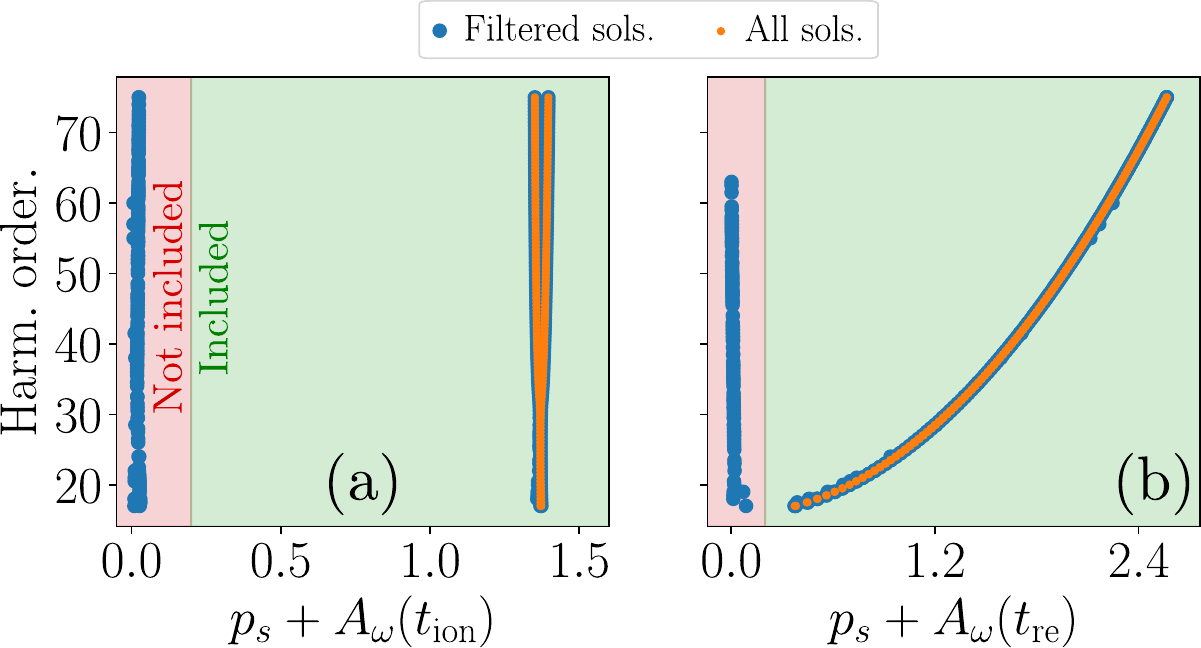}
	\caption{Panels (a) and (b) show the function $p+A(t)$ evaluated on the saddle-points when setting (a) $t = t_{\text{ion}}$ and (b) $t = t_{\text{re}}$. In orange, the saddle-point solutions when using Eqs.~\eqref{Eq:ionization}-\eqref{Eq:recombination}. In blue, the same saddle-points but filtered only on those instances where the partial derivatives of $\Phi$ in the saddle-point equations perturbatively modify the saddle-point equations. Here, we have set $E_{2\omega} = 10^{-2}\times E_{\omega}$, with $E_{\omega} = 0.053$ a.u., $\omega = 0.057$ a.u. and $I_p = 0.9$ a.u. corresponding to Helium atoms.}
	\label{Fig:poles:filtering}
\end{figure}

However, it is important to note that the modified action used to derive the saddle-point equations is a complex function with poles, specifically of first order. This necessitates caution when analyzing the saddle points, as the integration contour cannot be arbitrarily deformed~\cite{olga_simpleman}. This is because our function does not only have saddle-points, as guaranteed by the Cauchy-Riemann conditions when having holomorphic functions.~Figure~\ref{Fig:poles:filtering} explicitly shows all the saddle-point solutions (in blue) we find for different harmonic orders when solving the system of equations Eqs.~\eqref{Eq:ionization}-\eqref{Eq:recombination}, as a function of $p_s + A_\omega(t)$. Specifically, panel (a) and (b) show the results for the ionization and the recombination times, respectively.

As can be observed, we find two well-defined sets of saddle-points, highlighted by the red and green regions. In the red region, the points lie close to the poles of the $\Phi$ function, as they satisfy $p_s+A_\omega(t) \approx 0$. These are not true saddle-points but rather locations where the action diverges---a fact that we can be confirmed by evaluating the Hessian of the action with respect to the integration variables. Therefore, these points should be excluded from the saddle-point analysis. In the absence of Fano resonances, this exclusion is particularly straightforward: as the harmonic order increases, the actual saddle-points become clearly separated from the divergent contributions. Consequently, in our analysis, we apply a filtering criterion and retain the solutions within the green region. The resulting points after applying this filtering are highlighted in orange. 

\begin{figure}[hb!]
	\centering
	\includegraphics[width=0.6\textwidth]{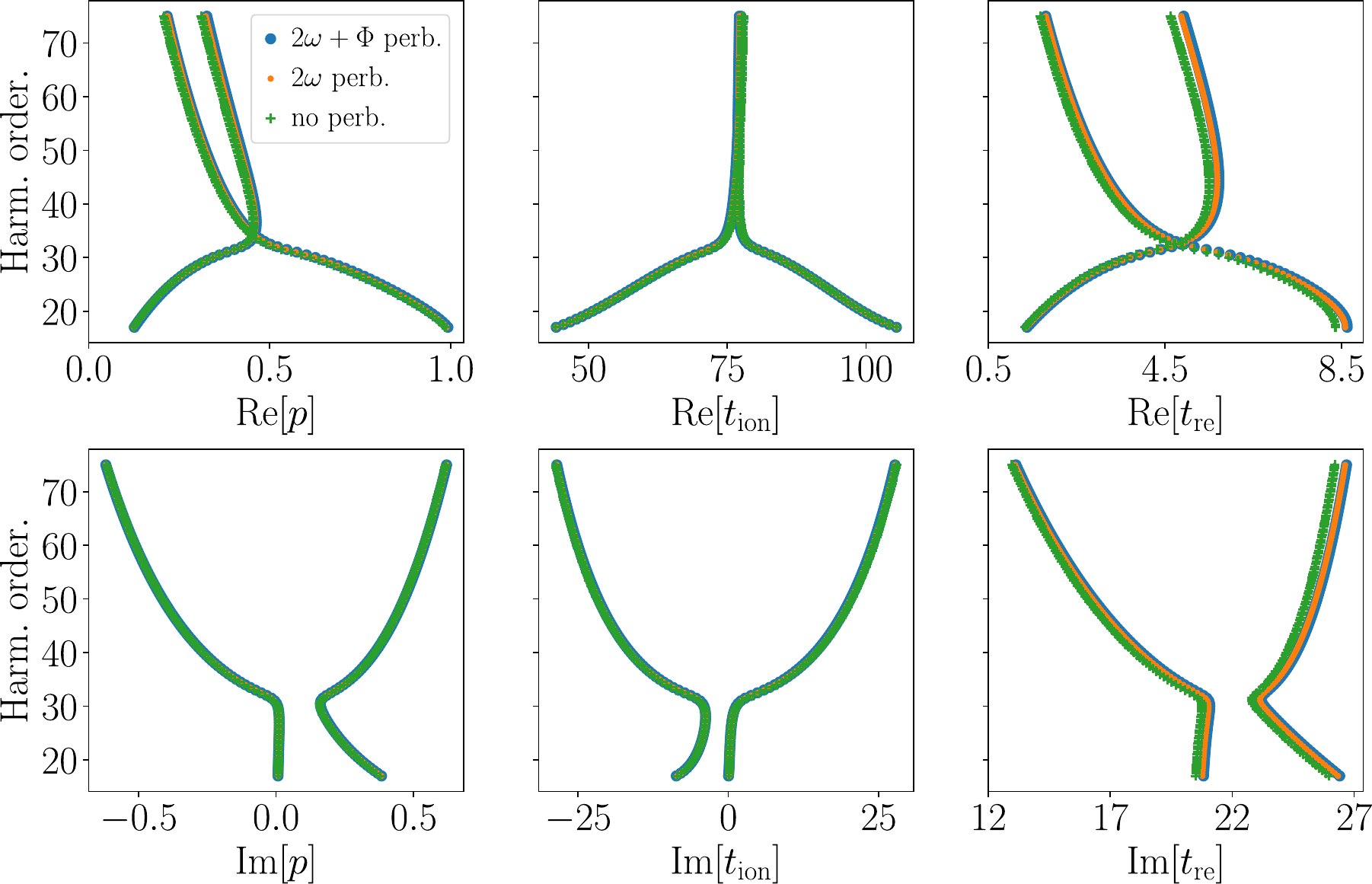}
	\caption{Saddle-point solutions when considering both the $\sigma$ and the $\Phi$ perturbations to the action (in blue), when considering just the $\sigma$ contribution (in orange) and when neither $\sigma$ and $\omega$ are included. Here, we have set $E_{2\omega} = 10^{-2} E_{\omega}$, with $E_{\omega} = 0.053$ a.u., $\omega = 0.057$ a.u. and $I_p = 0.9$ a.u. corresponding to Helium atoms.}
	\label{Fig:comparison:approx}
\end{figure}

In Fig.~\ref{Fig:comparison:approx} we plot the solutions to the saddle-point equations under three different scenarios: (i) when considering the action being modified by both $\sigma+ \Phi$, with the filtering described above applied and where $\sigma = \int \dd \tau [p+A_\omega(\tau)]A_{2\omega}(\tau)$ (blue); (ii) considering only the $\sigma$ perturbation (orange); and (iii) neglecting both perturbations (green). We observe that the addition of the $2\omega$ perturbs the saddle-points, particularly the recombination time, with the $\Phi$ contribution introducing a similarly perturbative effect.

\subsection{In the presence of the Fano resonance}
\begin{figure}
    \centering
    \includegraphics[width=0.8\textwidth]{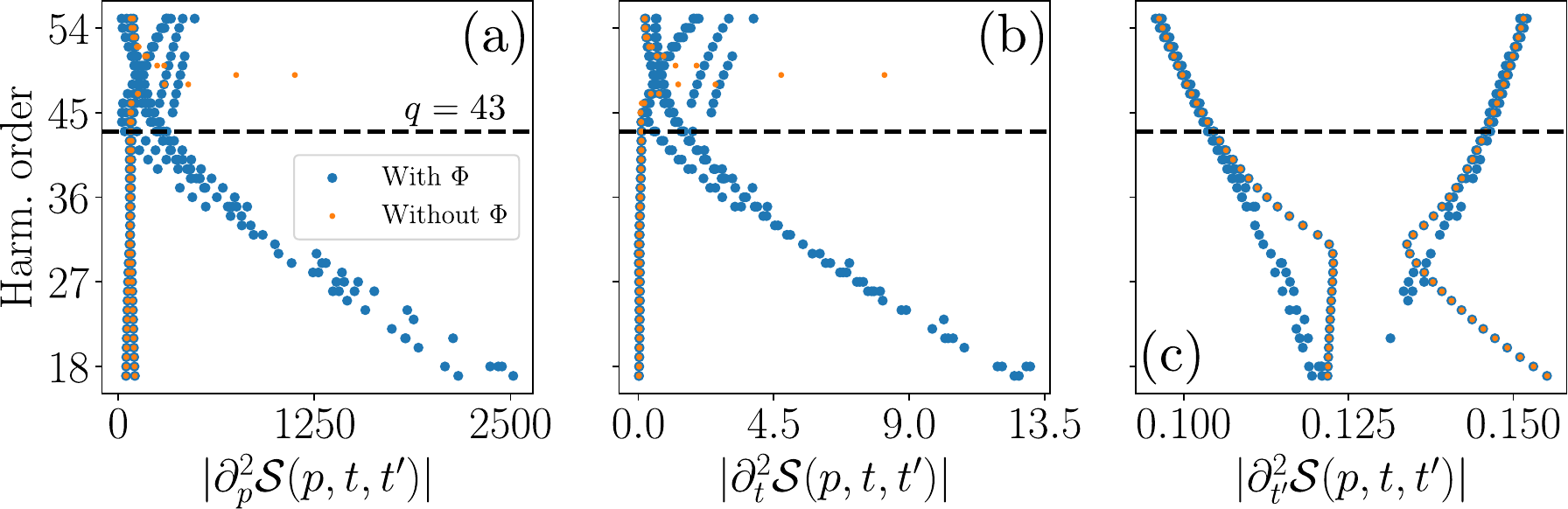}
    \caption{Hessian analysis for the saddle-points. Panels (a) to (c) present the absolute value of the diagonal elements of the Hessian with respect to the different integration variables. In blue markers represent the case where $\Phi$ is included while accounting for the Fano resonance on the saddle-point equations; in orange, when it is instead excluded.}
    \label{fig:hessian:Fano}
\end{figure}

A similar saddle-point analysis can, in principle, be carried out in the presence of the Fano resonance. As before, we find both genuine saddle-point solutions and divergent contributions arising from the resonance, which manifest as divergences in the the Hessian. This behavior is illustrated in Fig.~\ref{fig:hessian:Fano}~(a), where the divergent contributions are shown in blue, in contrast to the regular case without the Fano resonance shown in orange. For harmonics orders well below the resonance ($q \ll 43$ for the case studied in the main text), a filtering procedure similar to that presented in Fig.~\ref{Fig:poles:filtering} can be applied, as the two set of solutions remain well separated in the parameter space $\{p_s, t_{\text{re}}, t_{\text{ion}}\}$, However, near the resonance, the divergent and saddle-point contributions become indistinguishable in parameter space. Interestingly, we observe that the Hessian no longer diverges in this regime, suggesting that a saddle-point analysis excluding $\Phi(p,t,t',\phi)$ remains sufficient to understand how the spectrum is modified. 

\begin{figure}
    \centering
    \includegraphics[width=0.5\textwidth]{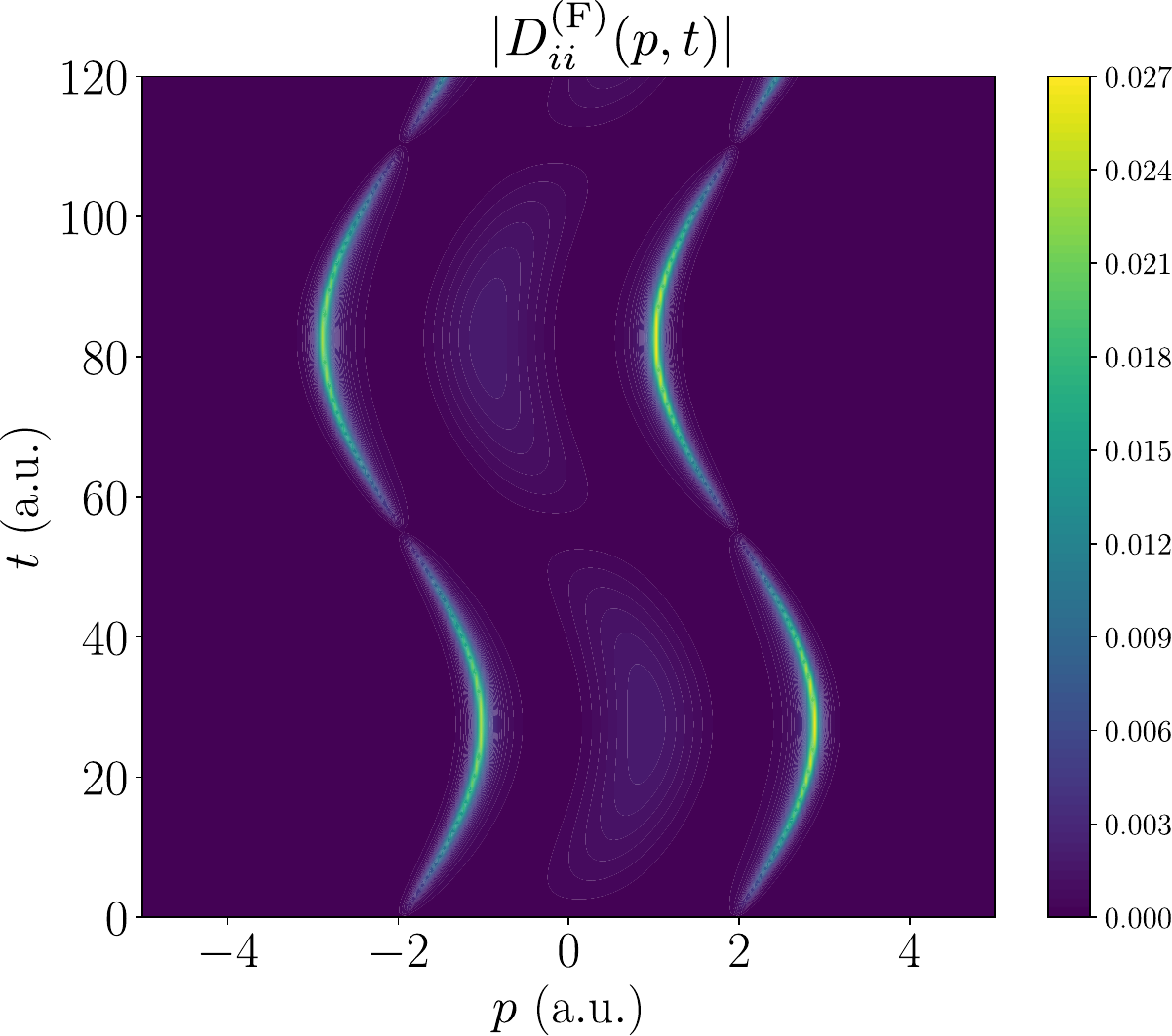}
    \caption{Surface of $\lvert D_{ii}^{(\text{F})}(p,t)\rvert$ evaluated over the $p$ and $t$, corresponding to the integration path used in the spectrum analysis.}
    \label{fig:Dii:surface}
\end{figure}

To further support this point, Fig.~\ref{fig:Dii:surface} shows a surface plot of $D_{ii}^{(\text{F})}(p,t)$ evaluated over real values of $p$ and $t$, which ultimately determine the integration path in the original dipole expression. We find that $|D_{ii}^{(\text{F})}(p,t)| \leq 0.03$, which is significantly smaller than the phase contribution of the unperturbed action contribution of the unperturbed action (on the order of $I_p = 0.9$ a.u.). Therefore, it does not significantly contribute to the rapidly oscillating behavior of $\mathcal{S}(p,t,t')$. As a result, when analyzing the spectrum, we use the unperturbed saddle-point solutions---that is, those obtained without including $\Phi(p,t,t',\phi)$.  

\section{Analysis when using squeezed $2\omega$ components}
Here, we investigate how the electron trajectories, the spectrum, and the properties of the outgoing light are modified when driving the system with squeezed $2\omega$ fields of the form
\begin{equation}
	\ket{\Phi(t)}
		= \ket{\bar{\alpha}_{\omega}} 
			\otimes
			\big[
				\hat{D}_{2\omega}(\bar{\alpha}_{2\omega})
					\hat{S}(\xi)
					\ket{0}
			\big],
\end{equation}
where $\alpha_{2\omega} = \epsilon \alpha_{\omega} e^{i\phi}$, with $\epsilon \ll 1$. Our analysis builds on the formalism introduced in Refs.~\cite{gorlach2020quantum,even_tzur_photon-statistics_2023,rivera2025structured}, and we therefore focus here only on the key equations relevant to our results.

\subsection{The HHG spectrum}
According to Refs.~\cite{gorlach2023high,rivera2025structured}, when using squeezed-light drivers, the HHG spectrum can be computed as
\begin{equation}
	S(\omega)
		= \dfrac{1}{\sqrt{2\pi\varsigma_i}}
			\int \dd \varepsilon^{(i)}_{2\omega}
				\exp[
					-\dfrac{\big(
									\varepsilon^{(i)}_{2\omega}
									 - \bar{\varepsilon}^{(i)}_{2\omega}
								 \big)^2}{2\varsigma_i}]
				\big\langle 
					d(\omega,\varepsilon^{(i)}_{2\omega})
				\big\rangle,
\end{equation}
where the superscript $i = x,y$ indicates the type of squeezing, and $\varsigma_i = 4 I_{\text{squ}}$ quantifies the squeezing strength.~In our case, the field components are given by $\bar{\varepsilon}_{2\omega}^{(x)} = \epsilon \bar{\varepsilon}_{\omega}\cos(\phi)$ and $\bar{\varepsilon}_{2\omega}^{(y)} = \epsilon \bar{\varepsilon}_{\omega}\sin(\phi)$, where $\bar{\varepsilon}_{\omega} \in \mathbbm{R}$.

\begin{figure}[h!]
	\centering
	\includegraphics[width=0.8\textwidth]{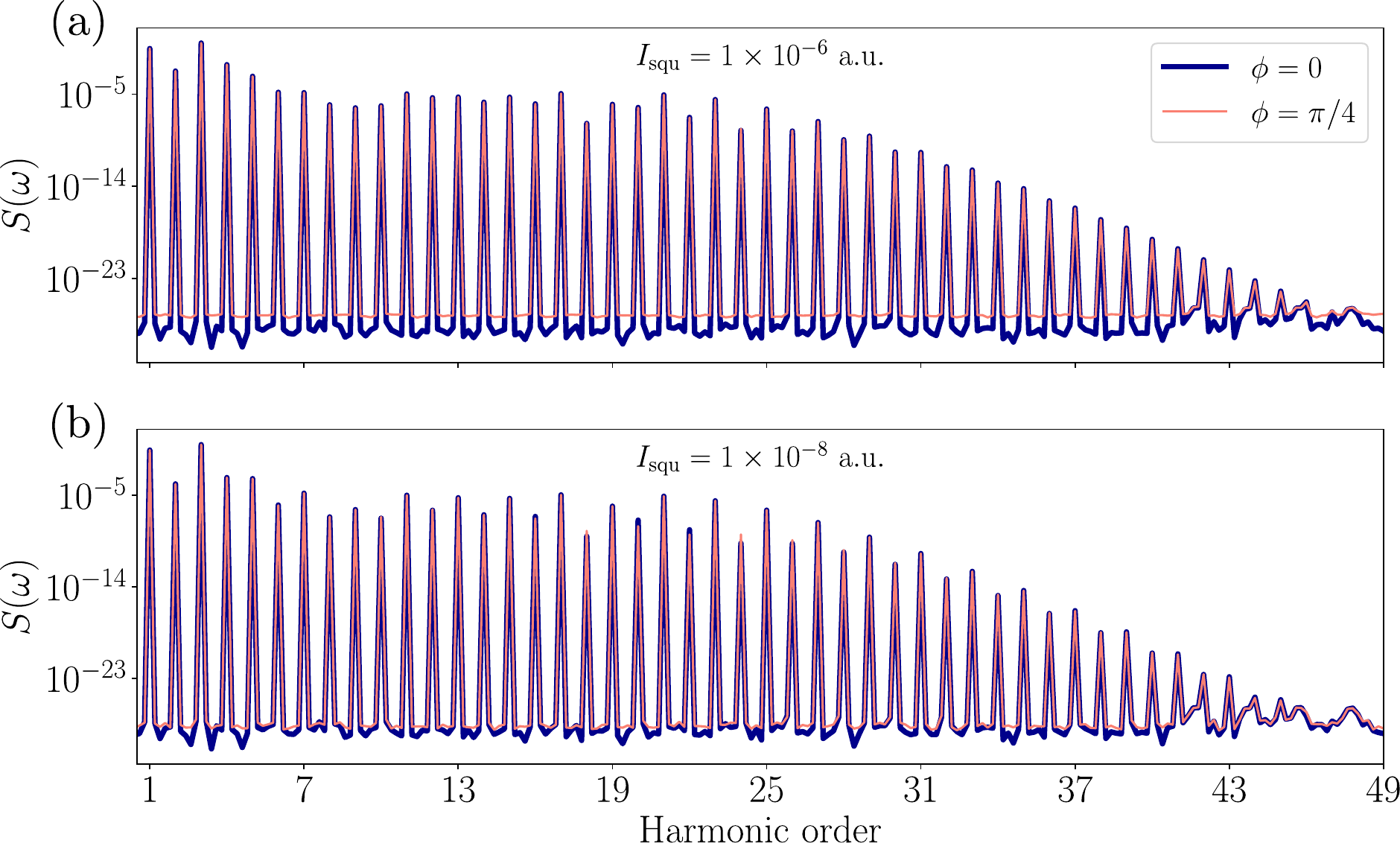}
	\caption{HHG spectra computed when considering the $2\omega$ field in a squeezed state with (a) $I_{\text{squ}} = 10^{-6}$ a.u., and (b) $I_{\text{squ}} = 10^{-8}$ a.u. and for different two-color phases.}
	\label{Fig:HHG:squ}
\end{figure}

In Figure~\ref{Fig:HHG:squ}, we present the computed HHG spectra for different levels of squeezing and various two-color delays. Increasing the amount of squeezing leads to more pronounced intensities in the even-order harmonics, while the HHG cutoff remains largely unchanged—consistent with expectations, as the squeezing does not alter the dominant $\omega$-frequency driver. In contrast, varying the two-color delay $\phi$ has little effect on the overall spectral structure.

\subsection{Quantum orbits}
As mentioned earlier, the introduction of squeezing can non-trivially affect the electron quantum orbits. To investigate this effect, we analyze the expectation value of the time-dependent dipole moment~\cite{even_tzur_photon-statistics_2023,rivera2025structured}, which is given by
\begin{equation}
	\langle \hat{d}(\Omega)\rangle
		= \dfrac{1}{\sqrt{8\pi I_{\text{squ}}}}
				\int \dd t_2 \int \dd\varepsilon_{2\omega}^{(i)}
					e^{i\Omega t_2}
					\exp[
							-\dfrac{
								\big(
									\varepsilon^{(i)}_{2\omega}
									- \bar{\varepsilon}^{(i)}_{2\omega}
								\big)^2}{2\varsigma_i}]
					\langle \phi(t_2,\varepsilon_{2\omega}^{(i)})\vert
						 \hat{d} 
					\vert \phi(t_2,\varepsilon_{2\omega}^{(i)})\rangle
\end{equation}
where $\relaxket{\phi(t,\varepsilon_{2\omega}^{(i)})}$ denotes the solution to
\begin{equation}
	i\hbar\pdv{\relaxket{\phi(t,\varepsilon_{2\omega}^{(i)})}}{t}
		= \Big[
				\hat{H}_{\text{at}}
				+ \mathsf{e} \hat{r} 
					\big(
						F_{\omega}(t) 
						+ F_{2\omega}(t,\varepsilon_{2\omega}^{(i)})
					\big)
			\Big]
				\relaxket{\phi(t,\varepsilon_{2\omega}^{(i)})},
\end{equation}
with $F_{2\omega}(t,\varepsilon_{2\omega}^{(i)}) =  \varepsilon_{2\omega}^{(y)}\cos(2\omega t) - \varepsilon_{2\omega}^{(x)}\sin(2\omega t)$, and where the superscript$i$ indicates the optical quadrature along which the squeezing is applied. In our case, we solve this equation within the SFA~\cite{lewenstein1994theory,amini2019symphony}.

Since the amplitude of one of the $2\omega$ field components enters as an integration variable, this case goes slightly beyond the analyses we have performed thus far. To account for this, we define the perturbation $\sigma(p,t,t')$ as
\begin{equation}
	\sigma(p,t_2,t_1)
		= 2\int^{t_2}_{t_1} \dd \tau
				\big[
					p + A_{\omega}(\tau)
				\big] A_{2\omega}(\tau,\varepsilon_{2\omega}^{(i)})
			+ \int^{t_2}_{t_1} \dd \tau
					A^2_{2\omega}(\tau,\varepsilon_{2\omega}^{(i)}),
\end{equation}
explicitly including the additional $A^2_{2\omega}(\tau,\varepsilon_{2\omega}^{(i)})$ term for completeness. This allows us to define a quantum-optical version of the action
\begin{equation}
	S^{(\text{QO})}(p,t_2,t_1)
		= \dfrac12 
			\int^{t_2}_{t_1} \dd \tau
				\Big[ 
					p + A_{\omega}(\tau) + 	A_{2\omega}(\tau,\varepsilon_{2\omega}^{(i)})
				\Big]^2
			+ I_p(t_2-t_1) -\Omega t_2
			- \dfrac{i}{2\sigma_i}
					\Big(
						\varepsilon^{(i)}_{2\omega}
						- \bar{\varepsilon}^{(i)}_{2\omega}
					\Big)^2.
\end{equation}
Applying the saddle-point approximation to evaluate all integrals, we obtain the following set of saddle-point equations
\begin{align}
	&\dfrac{
		\big[ 
				p_s + A_{\omega}(t_{\text{ion}})
				 + 	A_{2\omega}(t_{\text{ion}},\varepsilon_{2\omega,s}^{(i)})
		\big]^2}{2}
		+ I_p = 0,
	\\&
	\int^{t_{\text{re}}}_{t_{\text{ion}}}
		\dd \tau
			\Big[ 
				p_s + A_{\omega}(\tau)
				+ 	A_{2\omega}(\tau,\varepsilon_{2\omega,s}^{(i)})
			\Big]
			=0,
	\\& 
	\dfrac{\big[ 
		p_s + A_{\omega}(t_{\text{re}})
		+ 	A_{2\omega}(t_{\text{re}},\varepsilon_{2\omega,s}^{(i)})
	\big]^2}{2}
		+ I_p = \Omega,
	\\&
	\dfrac{1}{2\omega}
	\int^{t_{\text{re}}}_{t_{\text{ion}}}
		\dd \tau
		\Big[ 
			p_s + A_{\omega}(\tau)
			+ 	A_{2\omega}(\tau,\varepsilon_{2\omega,s}^{(i)})
		\Big]\cos(2\omega \tau)
			- \dfrac{i}{\sigma_i}
				\Big(
					\varepsilon^{(i)}_{2\omega,s}
					- \bar{\varepsilon}^{(i)}_{2\omega}
				\Big)
		= 0,
\end{align}
which we solve numerically, following the procedure outlined in previous sections.

\begin{figure}
	\centering
	\includegraphics[width=0.6\textwidth]{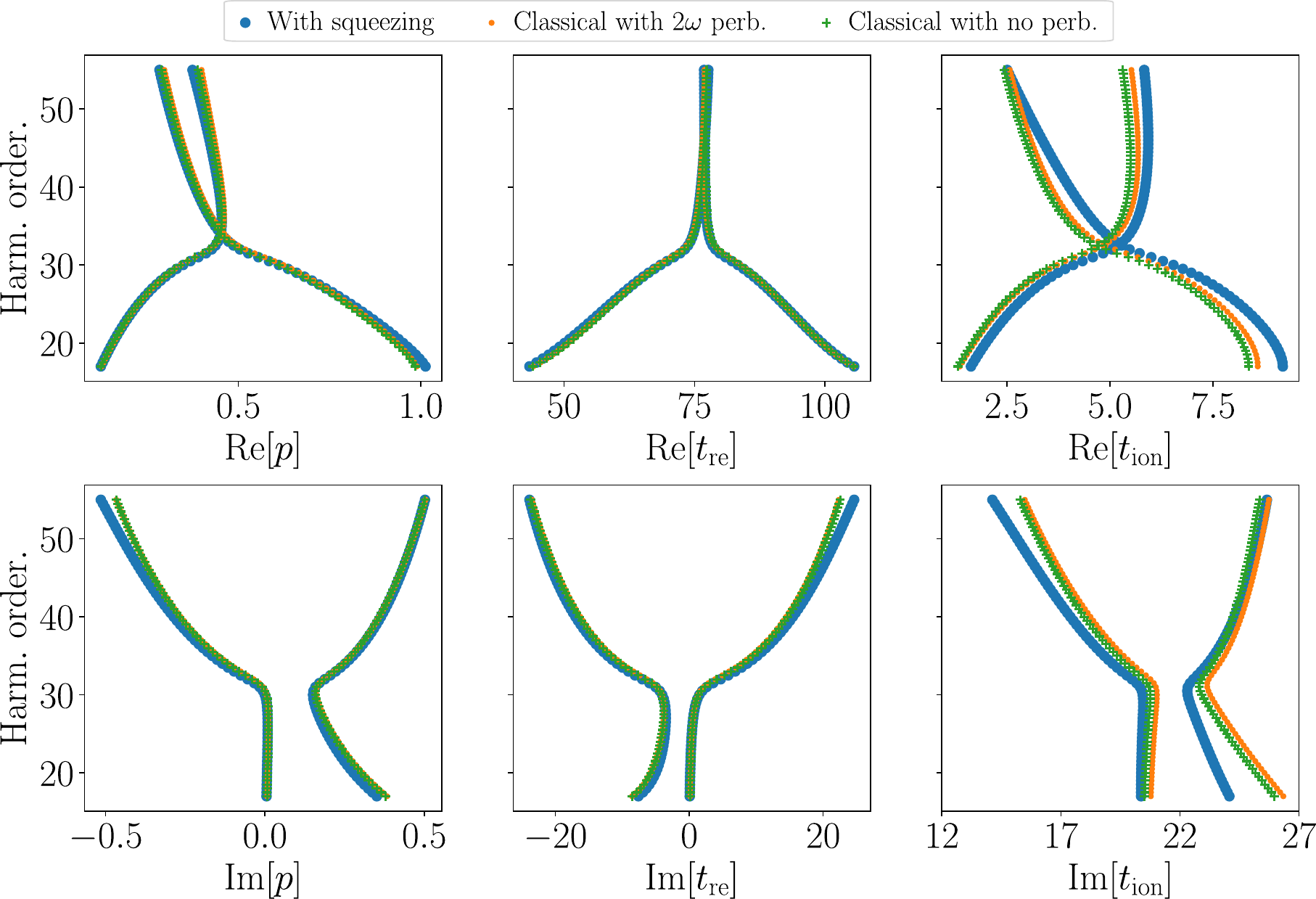}
	\caption{Saddle-point solutions when including squeezing (in blue), when considering classical fields and just the $\sigma$ contribution to the action (in orange) and when neither $\sigma$ nor squeezing are included. Here, we have set $I_{\text{squ}} = 1\times 10^{-6}$ a.u., $E_{2\omega} = 10^{-2}\times E_{\omega}$, with $E_{\omega} = 0.053$ a.u., $\omega = 0.057$ a.u. and $I_p = 0.9$ a.u. corresponding to Helium atoms.}
	\label{Fig:Trajectories:NC}
\end{figure}

Figure~\ref{Fig:Trajectories:NC} illustrates how the electron quantum orbits are modified in the presence of squeezing, comparing the case $I_{\text{squ}} = 10^{-6}$ a.u. (in blue) and $\phi = 0$, the reference cases without squeezing (orange) and without both squeezing and the $\sigma(p,t,t')$ perturbation. We observe that squeezing primarily affects the imaginary parts of the saddle-point solutions, as well as to the real part of the ionization time. Notably, increasing the squeezing intensity leads to the emergence of additional quantum orbits. However, in this work, we focus on the standard short and long trajectories, which remain dominant at $I_{\text{squ}}=10^{-6}$ a.u.---a value compatible with current state-of-the-art squeezing sources.

\subsection{Wigner functions}
According to Ref.~\cite{gorlach2023high}, after the light-matter interaction with an arbitrary driving field, the combined electron-light state can be expressed in terms of the generalized positive $P$-representation~\cite{drummond_generalised_1980}
\begin{equation}
	\hat{\rho}(t)
		= \int \dd^2 \alpha \int\dd^2 \beta\
			\dfrac{P(\alpha,\beta^*)}{\braket{\beta^*}{\alpha}}
				\dyad{\phi_{\alpha}(t)}{\phi_{\beta^*}(t)}
				\hat{D}(\alpha)
					\dyad{\chi_{\alpha,1}(t)}{\chi_{\beta^*,1}(t)}
					\hat{D}^\dagger(\beta^*)
				\bigotimes_{q\neq 1}
				\dyad{\chi_{\alpha,q}(t)}{\chi_{\beta^*,q}(t)},~
\end{equation}
where $P(\alpha,\beta^*)$ is a real-valued function describing the properties of the driving field. This expression can be in principle used as it is for computing different quantum optical properties of the outgoing light, in particular the Wigner function of the outgoing light. However, before doing so, we introduce some approximations that ease the calculations of these.

Firstly, following Ref.~\cite{tzur2024generation}, we approximate
\begin{equation}
	\braket{\boldsymbol{\chi}_{\beta^*}}{\boldsymbol{\chi}_{\alpha}}
	\braket{\phi_{\beta^*}(t)}{\phi_\alpha(t)}
	\approx \braket{\beta^*}{\alpha},
\end{equation}
where we have denoted, for simplicity, $\ket{\boldsymbol{\chi}(t)} \equiv \bigotimes_{q} \ket{\chi_{q}(t)}$. This approximation is valid under the assumption that the electron backaction during the propagation step does not significantly influence the quantum state of the harmonics~\cite{rivera2022light,rivera-dean_role_2024}. Consequently, we can write the reduced density matrix of the field as
\begin{equation}
	\hat{\rho}_f(t)
	= \int \dd^2 \alpha \int \dd^2\beta
	\dfrac{P(\alpha,\beta^*)}{\braket{\boldsymbol{\chi}_{\beta^*}(t)}{\boldsymbol{\chi}_{\alpha}(t)}}
	\dyad{\boldsymbol{\chi}_{\alpha}(t)}{\boldsymbol{\chi}_{\beta^*}(t)},
\end{equation}
and, for a single harmonic mode,
\begin{equation}\label{Eq:state:harmonics}
	\hat{\rho}_{q}(t)
	= \int \dd^2 \alpha \int \dd^2\beta
	\dfrac{P(\alpha,\beta^*)}{\braket{\chi_{\beta^*,q}(t)}{\chi_{\alpha,q}(t)}}
	\dyad{\boldsymbol{\chi}_{\alpha,q}(t)}{\boldsymbol{\chi}_{\beta^*,q}(t)},
\end{equation}
which serves as the basis for our further calculations.

For analyzing properties involving the harmonic modes, it becomes particularly beneficial to work both in the classical limit and the thermodynamic limit. These two limits are defined as follows:
\begin{itemize}
	\item \textbf{Classical limit.} In this limit, the coherent state amplitude is expressed as $\alpha = 2 \epsilon \varepsilon_\alpha$, where $\varepsilon_\alpha$ represents the electric field amplitude. The classical limit is achieved by letting the quantization volume become arbitrarily large, i.e., $V \to \infty$, which corresponds to $\epsilon \to 0$.
	
	\item \textbf{Thermodynamic limit.} This limit is particularly relevant for analyzing the harmonic modes, whose coherent state amplitude is given by $\chi_{\alpha,q}(t) = \epsilon d(\omega_q)$, where $d(\omega_q)$ denotes the Fourier transform of the time-dependent dipole moment. In the classical limit, as $\epsilon \to 0$, we find that $\chi_{\alpha,q}(t) \to 0$. This can be interpreted as a consequence of the coupling between light and matter becoming negligibly small in an infinitely large laser-matter interaction region, thus making the probability of generating harmonic radiation vanishingly small.
	
	However, in typical HHG experiments, the harmonic radiation arises from the collective contribution of many atoms. As described in Refs.~\cite{lewenstein2021generation,rivera2022strong,stammer2023quantum}, the phase-matched contribution from $N$ atoms leads to a coherent enhancement, with $\chi_{\alpha,q} \to N \chi_{\alpha,q}$. To account for this, we introduce the thermodynamic limit: when taking the limit $V \to \infty$ (equivalently $\epsilon \to 0$), we assume that $N\to \infty$ such that $N\epsilon = \kappa= \text{constant}$. 
\end{itemize}

These limits significantly simplify the analysis of quantum optical properties of the harmonics. For example, let us consider a linear operation $\Lambda_{q}[\cdot]$ that, when applied to Eq.~\eqref{Eq:state:harmonics}, results in
\begin{equation}
	\Lambda_{q}
	\big[
	\hat{\rho}_{q}(t)
	\big]
	= \int \dd^2 \alpha \int \dd^2 \beta
	\dfrac{P(\alpha,\beta^*)}{\braket{\chi_{\beta^*,q}(t)}{\chi_{\alpha,q}(t)}}
	f\big(
	\chi_{\beta^*,q}(t)),
	\chi_{\alpha,q}(t)
	\big),
\end{equation}
where we consider that the function $f(\cdot)$ does not introduce additional dependencies neither with $\epsilon$ nor $N$. In such instance, the use of the classical and thermodynamic limits depicted above yield
\begin{equation}
	\lim_{\epsilon\to 0\ \text{s.t.}\ N\epsilon \to \kappa}
	\Lambda_{q}
	\big[
	\hat{\rho}_{q}(t)
	\big]
	= \int \dd^2 \varepsilon_\alpha\int \dd^2 \varepsilon_\beta
	\bigg[
	\lim_{\epsilon \to 0}
	\dfrac{1}{16\epsilon^4}P(\varepsilon_\alpha,\varepsilon_\beta^*)
	\bigg]
	\dfrac{f\big(\kappa d_{\varepsilon_\alpha}(\omega_{q}),\kappa d_{\varepsilon_{\beta^*}}(\omega_{q})\big)
	}{\braket{\kappa d_{\varepsilon_{\beta^*}}(\omega_{q})}{\kappa d_{\varepsilon_\alpha}(\omega_{q})}}.
\end{equation}

\begin{figure}
    \centering
    \includegraphics[width=0.8\textwidth]{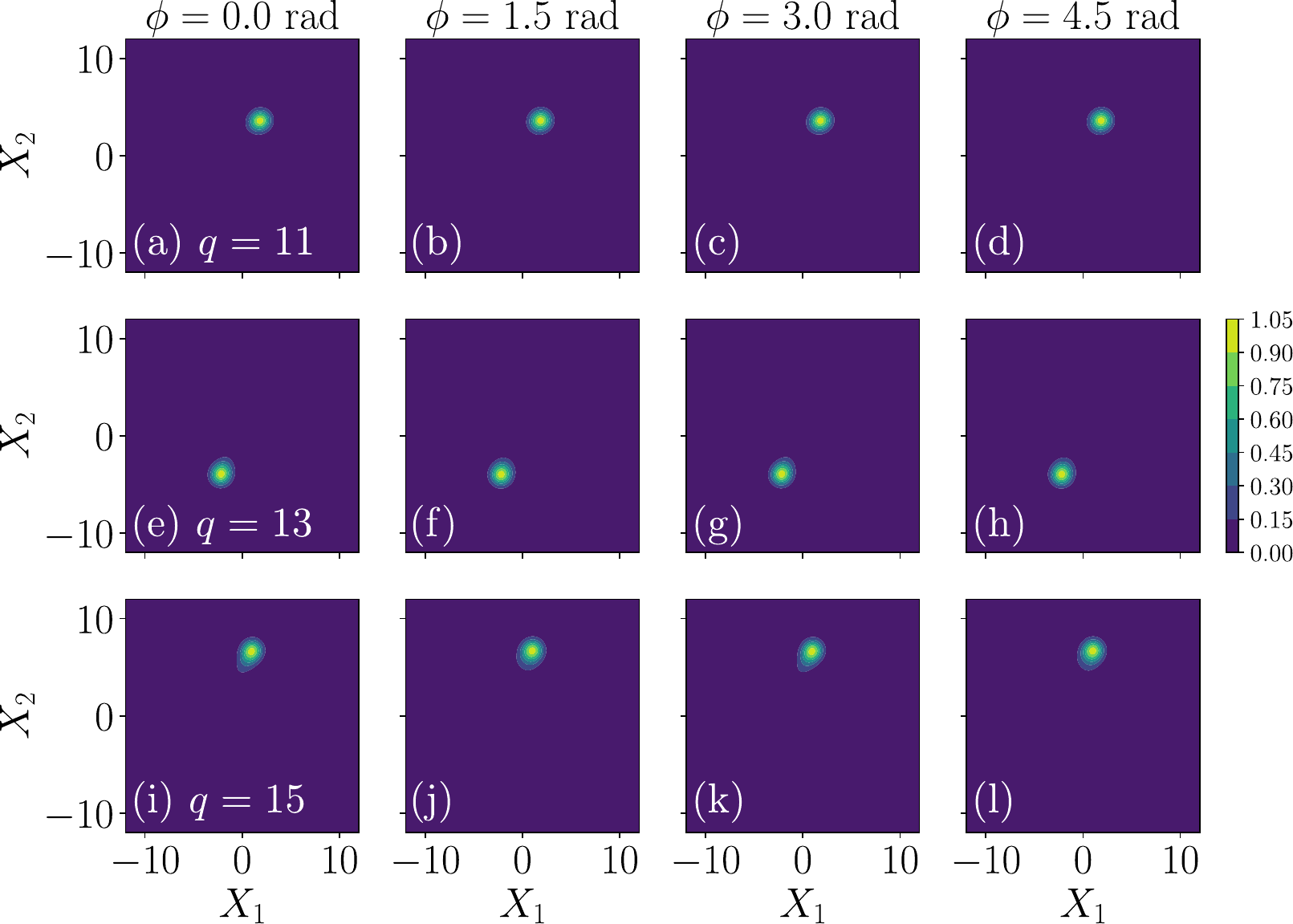}
    \caption{Wigner function for different odd harmonic orders as a function of the two-color delay. Here, we set $\kappa = 2.5\times 10^6$ (with the value depending on the number of points used for computing the Fourier Transform) and $I_{\text{squ}} = 10^{-6}$ a.u.~for the squeezing intensity.}
    \label{Fig:Wigner:odd}
\end{figure}

An example of such linear operation is the Wigner function, characterized by $\mathcal{W}_\gamma[\hat{\rho}] = \text{tr}(\hat{D}(\gamma) \hat{\Pi}\hat{D}(-\gamma)\hat{\rho})$~\cite{royer_wigner_1977}. In this particular case, we find
\begin{equation}
	W_{q}(\gamma)
	= \dfrac{1}{\sqrt{2\pi \varsigma_i}}
	\int \dd \varepsilon_{\alpha,i}
	\exp[-\dfrac{\big(\varepsilon_{\alpha,i}-\bar{\varepsilon}_i\big)^2}{2\varsigma_i}]
	\langle \kappa d_{\varepsilon_{\alpha}}(\omega_{q})\vert 
	\hat{D}(\gamma)\hat{\Pi}\hat{D}(-\gamma)
	\vert \kappa d_{\varepsilon_{\alpha}}(\omega_{q})\rangle,
\end{equation}
where we remind that $\varepsilon_{\alpha} = \varepsilon_{\alpha,x} + i \varepsilon_{\alpha,y}$, and the integral is done over either the real or the imaginary part of this quantity. This constitutes the central equation we use in computing the Wigner functions shown in the main text. In our numerical calculations we set $\kappa = 2.5 \times 10^6$, though this value crucially depends on the number of points used to compute the Fourier Transform that results in $d_{\varepsilon_{\alpha}}(\omega_{q})$.

As a complementary plot to those presented in the main text, Fig.~\ref{Fig:Wigner:odd} shows the Wigner function for different odd harmonic orders as a function of the two-color delay $\phi$.  In all cases the resulting state exhibits a Gaussian-like Wigner function, albeit with slightly super-Poissonian characteristics, as indicated by the $g^{(2)}(\tau=0)$ being slightly greater than one. This behavior is most evident for 15th harmonic orders [panels (i)-(l)], where a faint tail pointing toward the origin can be observed. Notably, the overall shape of the Wigner function appears largely insensitive to variations in the two-color delay.

\end{document}